\documentclass[10pt,journal,compsoc]{IEEEtran}
\ifCLASSOPTIONcompsoc
  \usepackage[nocompress]{cite}
\else
  \usepackage{cite}
\fi
\ifCLASSINFOpdf
\else
\fi
\usepackage{cite}
\usepackage{amsmath,amssymb,amsfonts}
\usepackage{graphicx}
\usepackage{listings}
\usepackage{textcomp}
\usepackage{xcolor}
\usepackage{subfigure}
\usepackage{algcompatible}
\usepackage{threeparttable}
\usepackage{tikz}
\usetikzlibrary{fit, calc}
\usetikzlibrary{decorations}
\usepackage{verbatim}
\usepackage{pgfplotstable}
\usepackage{diagbox}
\usepackage{makebox}
\usepackage{multirow}
\usepackage{makecell}
\usepackage{booktabs}
\usepackage{epsfig}
\usepackage[numbers,sort&compress]{natbib} 
\usepackage[justification=centering]{caption}
\usepackage{float}
\usepackage{graphicx}

\usepackage{algorithm}%
\usepackage{algorithmicx}%
\usepackage{algpseudocode}%
\usepackage{subcaption}

\usepackage{multirow}

\usepackage{ listings}
\usepackage{xcolor}
\definecolor{mygreen}{rgb}{0,0.6,0}
\definecolor{mygray}{rgb}{0.5,0.5,0.5}
\definecolor{mymauve}{rgb}{0.58,0,0.82}
\begin{document}
\title{A Practical Adversarial Attack against Sequence-based Deep Learning Malware Classifiers}


\author{
        Kai Tan,
        Dongyang Zhan*,~\IEEEmembership{Member,~IEEE,} 
        Lin Ye,
                      %~\IEEEmembership{Fellow,~OSA,}
        Hongli Zhang, %,~\IEEEmembership{Life~Fellow,~IEEE}% <-this % stops a space
        and~Binxing Fang                         
\IEEEcompsocitemizethanks{\IEEEcompsocthanksitem K. Tan, D. Zhan, L. Ye, H. Zhang, B Fang. are with the School of Cyberspace Science, Harbin Institute of Technology, Harbin,
Heilongjiang, 150001.
% \protect\\
E-mail: 20b903076@stu.hit.edu.cn, \{zhandy, hityelin, zhanghongli, fangbx\}@hit.edu.cn 

%\IEEEcompsocthanksitem J. Doe and J. Doe are with Anonymous University.
\IEEEcompsocthanksitem * Corresponding Author: zhandy@hit.edu.cn}% <-this % stops an unwanted space
%}% <-this % stops an unwanted space
}

\markboth{Journal of \LaTeX\ Class Files,~Vol.~14, No.~8, August~2015}%
{Shell \MakeLowercase{\textit{et al.}}: Bare Demo of IEEEtran.cls for Computer Society Journals}



\IEEEtitleabstractindextext{%
\begin{abstract}
Sequence-based deep learning models (e.g., RNNs), can detect malware by analyzing its behavioral sequences. Meanwhile, these models are susceptible to adversarial attacks. Attackers can create adversarial samples that alter the sequence characteristics of behavior sequences to deceive malware classifiers. The existing methods for generating adversarial samples typically involve deleting or replacing crucial behaviors in the original data sequences, or inserting benign behaviors that may violate the behavior constraints. However, these methods that directly manipulate sequences make adversarial samples difficult to implement or apply in practice. In this paper, we propose an adversarial attack approach based on Deep Q-Network and a heuristic backtracking search strategy, which can generate perturbation sequences that satisfy practical conditions for successful attacks. Subsequently, we utilize a novel transformation approach that maps modifications back to the source code, thereby avoiding the need to directly modify the behavior log sequences. We conduct an evaluation of our approach, and the results confirm its effectiveness in generating adversarial samples from real-world malware behavior sequences, which have a high success rate in evading anomaly detection models. Furthermore, our approach is practical and can generate adversarial samples while maintaining the functionality of the modified software.

\end{abstract}

}

\maketitle

\IEEEdisplaynontitleabstractindextext

% \fi
\IEEEpeerreviewmaketitle

\IEEEraisesectionheading{\section{Introduction}\label{sec:introduction}}


 \IEEEPARstart{I}{n} recent years, the field of adversarial deep learning has attracted significant attention, driven by the need to understand the vulnerabilities of deep learning models and devise methods to exploit them. Adversarial attacks on sequence-based classifiers, particularly those employed for detecting malicious sequences generated by malware, have become a critical research topic \cite{grosse2016adversarial,papernot2017practical,peng2021semantics,yan2022survey}. The primary objective of these attacks is to generate adversarial behavior sequences that can evade anomaly detection systems \cite{luo2020novel}, while preserving their intended malicious behaviors.

The adversarial attack techniques for behavior sequences primarily rely on direct modification of the data, such as manipulating behavior log sequences to evade detection by deep learning models. Although these attack methods \cite{herath2021real,lu2023black} may demonstrate effectiveness in experimental settings, they can be impractical and face challenges when used in real-world scenarios. In typical scenarios, anomaly detection models process log data quickly, while intercepting and modifying behavior logs before processing requires sufficiently high system privileges. In addition, even if an attacker can directly tamper with behavior logs, it is also necessary to consider the situation where logs are encrypted or protected by other security mechanisms. Consequently, this assumption may encounter many technical and operational challenges during practical adversarial attacks.

% For instance, attackers might try to delete or replace malicious system calls or APIs of the malware to mislead the classifiers. \textcolor{red}{However, such modifications can be challenging to implement and may also impact the intended functionality of the program.
% Since System calls and APIs often have inter-dependencies, these operations could introduce errors or incompatibilities.}

%as attackers need to modify the original program in order to produce effective adversarial sequences. 
% For instance, attackers might try to delete or replace malicious system calls or APIs of the malware to mislead the classifiers. However, such modifications can be challenging to implement and may also impact the intended functionality of the program. Attackers require deep understanding of the program codes and must be familiar with the implementation details of the program code, in order to determine which behaviors can be modified without affecting the program. As a result, these attack methods demand significant technical and resource investment, along with professional knowledge base.

% In addition to directly modifying data, 
Some other adversarial attack methods \cite{hu2017black,rosenberg2018generic,rosenberg2020query} have been developed to consider practicality and feasibility, such as inserting benign behaviors to mislead the deep learning model. These methods do not directly modify behavior logs. Instead, they modify the malware and indirectly alter the behavioral sequences. In addition, unlike replacing or deleting malicious system calls or APIs, these methods only insert benign behaviors into malicious sequences to preserve the core functionality of the malware.

%However, attackers need to deeply understand the structure and implementation details of the original program code to determine where to insert additional behaviors, system calls, or APIs to mislead the deep learning model. 

% For instance, consider a malicious system call sequence: \{\texttt{exec, fork connect, setuid, exit}\}. By inserting benign syscalls like "\texttt{open}", "\texttt{read}", and "\texttt{close}" at appropriate positions in this sequence, the attacker could potentially mislead the classifier identifying the modified sequence as benign.

However, some of these methods also face various challenges. First, the attackers cannot insert numerous benign behaviors into the sequence, which could make the adversarial sequences more detectable and raise suspicions. Effectively achieving minor perturbations presents a notable challenge, as it requires a strategic approach to determining the optimal positions and types of benign behaviors for insertion within the sequence. Secondly, inserting benign behaviors into a malicious sequence without meeting constraints may result in an invalid or semantically incorrect sequence. For example, inserting the benign behaviors "close\_file" and "read\_file" consecutively in a malicious sequence can cause errors if the file is closed before it is read, resulting in an invalid sequence. Moreover, these methods have no specific guidance on how to modify the attack source code to achieve the desired insertion of behaviors. Attackers also need to deeply understand the structure and implementation details of the original program code and map modifications back to the source code.

To address the above challenges, we propose a practical approach for generating adversarial samples against sequence-based deep learning malware classifiers. We also consider the less difficult way to insert benign behavior into the sequence to achieve adversarial attacks. 

Specifically, we first construct a perturbation action set consisting of multiple groups of related benign behaviors, which implies the range of possible perturbations. In addition, we strategically select highly representative benign behavior fragments to enhance the quality of adversarial samples.

Subsequently, we train a Deep Q-Network (DQN) model \cite{mnih2015human} to learn the optimal behaviors to modify the sequence, maximizing the perturbation effect. In comparison to other methods, our approach can dynamically select the most suitable perturbation behaviors based on the current sequence and minimize the extent of modifications required.

% \textcolor{red}{In comparison to other methods, our utilization of a Deep Q-Network (DQN) model offers distinct advantages. By dynamically selecting the most suitable perturbation actions based on the current sequence, our approach minimizes the extent of modifications required. This contrasts with traditional methods that might involve more extensive alterations, leading to potential loss of original functionality. As a result, our DQN-based strategy not only enhances the perturbation effect but also maintains the core characteristics of the original sequence more effectively than many alternative approaches.}

As previously mentioned, inserting benign behaviors can damage the semantic validity of the sequence. We then apply a heuristic backtracking search strategy based on the DQN model and the perturbation action set to meet these constraints of real-world sequence modifications.

% As The semantic validity of the sequence.
% We then apply a heuristic backtracking search strategy based on the DQN model and the perturbation action set to efficiently search for the optimal perturbation sequences.

% that can evade classifier detection while maintaining practicality. 

% \textcolor{red}{This approach gains its advantage over alternative methods by effectively addressing the constraints of real-world sequence modifications through heuristic search. This not only ensures the maintenance of practicality but also enhances the search efficiency. By efficiently navigating the search space, our method is uniquely suited to produce optimal perturbation sequences that can elude classifier detection while still adhering to real-world feasibility.}

Finally, due to the lack of specific guidance in most methods on how to modify attack source code to achieve the desired insertion behaviors, we design a transformation approach that maps modifications back to the source code. Our approach enables the correlation between the source code and its corresponding behavioral sequences, enabling automatic modifications at desired code locations.

% To illustrate the feasibility of attacks in real-world scenarios, we employ a transformation approach that maps modifications back to the source code, preservating the original functionality of the program.

%被动句
% To summarize, the key contributions are outlined as follows:
% \begin{itemize}
% \item A perturbation action set consisting of multiple groups of related benign behaviors is constructed, which can be used to generate adversarial sequences.
% % \item Optimal behaviors for modifying the sequence are learned using a Deep Q-Network (DQN) model, maximizing the perturbation effect while ensuring the preservation of the original program functionality.
% \item A heuristic backtracking search strategy, based on DQN and the perturbation action set, is implemented for efficiently identifying the optimal perturbation sequences that can evade classifier detection and maintain practicality in real-world scenarios.
% \item A transformation approach is introduced, which maps modifications back to the source code, preserving the original functionality of the program and providing guidance on how to modify the program code to achieve the desired behavior modification.
% \item A comprehensive experimental evaluation of our proposed approach, indicating its effectiveness in generating adversarial sequences that can bypass various classifiers. It also highlights the potential value of our approach in real-world applications.
% \end{itemize}

% 被动句
To summarize, our key contributions are outlined as follows:
\begin{itemize}
\item A practical adversarial attack approach against sequence-based deep learning malware classifiers is proposed.
\item A novel transformation approach is introduced, which maps sequence modifications back to the source code.
\item Comprehensive experiments are conducted after implementing the prototype of our approach.
\end{itemize}

% To summarize, the first contribution of this paper is that  a practical adversarial attack approach against sequence-based deep learning malware classifiers is proposed. It involves the construction of a perturbation action set comprising multiple groups of related benign behaviors \textcolor{red}{to generate adversarial sequences. } Additionally, a heuristic backtracking search strategy, based on DQN and the perturbation action set, is implemented to efficiently identify optimal perturbation sequences that can evade classifier detection while remaining practical in real-world scenarios.

% The second contribution is that a transformation approach is introduced, which maps modifications back to the source code, preserving the original functionality of the program and providing guidance on how to modify the program code to achieve the desired behavior modification.

% The last contribution is that after implementing the prototype, a comprehensive experimental is conducted to indicate the effectiveness and practicability in generating adversarial sequences that can bypass various classifiers.

The remaining sections of the paper are structured as follows: The background is introduced in Section \ref{s:Threat}. Section \ref{s:design} describes the overall framework and its implementation. The evaluation of our approach is performed in Section \ref{s:evaluation}. Section \ref{s:related-work} summarizes the related work. Section \ref{s:Discussion} and Section \ref{s:conclusion} respectively discuss and conclude this paper.

% Our work builds upon these recent advancements in adversarial attacks on sequence-based classifiers and proposes a novel method that combines deep reinforcement learning with behavior generation templates constructed from benign sequences. This approach allows for the generation of semantically valid adversarial sequences while maintaining the logical relationships among the behaviors, making it even more effective in deceiving machine learning models and evading security mechanisms.

% \subsection{Deep Learning-based Malware Detection}

% \subsection{Adversarial Attacks of Sequential Models}

% \subsection{Adversarial Defense Approaches}

\section{Background}\label{s:Threat}
In this section, we introduce the problem statement, the threat model of our approach, and the perturbation examples analysis.

\subsection{Problem Statement}

The problem we aim to solve involves generating an adversarial sequence $S'$, from an input sequence $S = \{s_1, s_2, \ldots, s_n\}$, which represents a series of behaviors or events of malware. The adversarial sequence should be capable of evading detection by a target classifier $C$ (e.g., an LSTM-based classifier), while maintaining its malicious functionality and minimizing the perturbation between the original sequence $S$ and the adversarial sequence $S'$. The perturbation can be quantified using the perturbation metric $d(S, S')$, which measures the ratio of the number of perturbed (inserted) elements in the sequence to the total length of the original sequence. Our goal is to find the optimal adversarial sequence $S'$ that minimizes the perturbation from the original sequence $S$. Furthermore, we hope to affect the behavior sequences by modifying the malware. Therefore, the generated adversarial sequence should be practically applicable, guiding the modification of the software source code without directly altering the behavior sequences. This allows us to create corresponding malicious software and verify attacks in practical scenarios.

\subsection{Threat Model}

To comprehensively understand the performance of a detection model, it is essential to investigate either its internal architecture of model or analyze the detection results based on the input provided. However, obtaining information about the model architecture can be difficult due to stringent security measures. 
Since adversarial samples demonstrate effectiveness across different models, attackers commonly employ black-box attacks \cite{rosenberg2018generic}. They can obtain the detection results by submitting specific inputs and creating a surrogate model to effectively emulate the target model, leading them to execute adversarial attacks based on the surrogate model. Before implementing the attack, attackers might engage in the systematic collection and analysis of substantial volumes of data through multiple low-level interactions with the target system. Additionally, the utilization of openly available datasets could serve as a resource to train effective surrogate models.
We assume that the attacker has not obtained high system privileges at the beginning of the attack, and the attacker has the source code of the malware so that he can modify the malware without directly manipulating the behavior log data.

% This makes it easier for the attackers to understand the behavior of the programs and modify the source codes to generate adversarial samples.

\subsection{Perturbation Examples Analysis}
We present an example to illustrate the difference between direct log data manipulation methods and our practical perturbation approach.

Suppose we have a malicious system call sequence: \{clone, execve, setuid, exit\_group\}. The corresponding malicious code is in Fig.\ref{fig:fig1}. Direct log data manipulation methods might modify the sequence by deleting or replacing certain system calls, such as changing the sequence to \{clone, execve, close, exit\_group\}. However, this could require higher system privileges and lead to unintended consequences, such as producing semantically incorrect sequences.

% Our practical perturbation approach:
Instead of directly manipulating the data, we insert benign system calls like openat, read, and close at appropriate positions in the sequence, for example: \{clone, openat, execve, read, close, setuid, exit\_group\}. Therefore, our approach can make the malicious sequence appear more benign to the classifier while not damaging the original functionality of the malware. This demonstrates the practicality and effectiveness of our perturbation approach in evading sequence-based deep learning malware classifiers.

In addition, our approach is also designed to map sequence modifications back to the source code. Suppose our perturbation sequence is \{clone, openat, execve, read, close, setuid, exit\_group\}. To match the modification, we need to insert codes related to openat, read, and close system calls at the specific positions, as shown in Fig\ref{fig:fig2}.

\begin{figure}[h]
\centering
\begin{lstlisting}[language={[ANSI]C},numbers=left, numberstyle=\tiny,keywordstyle=\color{blue!70},commentstyle=\color{red!50!green!50!blue!50},frame=single, rulesepcolor=\color{red!20!green!20!blue!20}]
#include <stdio.h>
#include <stdlib.h>
#include <unistd.h>
#include <sys/types.h>
#include <sys/wait.h>

int main() {
    pid_t pid = fork();
    if (pid >= 0) {
         execv("malware_program", NULL);}
    setuid(0);
    exit(0);}
\end{lstlisting}
\caption{An example of a malicious code.}
\label{fig:fig1}
\end{figure}

\begin{figure}[h]
\centering
\begin{lstlisting}[language={[ANSI]C},numbers=left, numberstyle=\tiny,keywordstyle=\color{blue!70},commentstyle=\color{red!50!green!50!blue!50},frame=single, rulesepcolor=\color{red!20!green!20!blue!20}, escapeinside={<@}{@>}]
#include <stdio.h>
#include <stdlib.h>
#include <unistd.h>
#include <sys/types.h>
#include <sys/wait.h>
#include <fcntl.h>
#include <sys/stat.h>

int main() {
    pid_t pid = fork();
    // Insert open syscall
<@\textcolor{red}{+  ~int fd = open("dummy\_file", O\_RDONLY);}@>
    if (pid >= 0) {
         execv("malware_program", NULL);}
<@\textcolor{red}{+  ~char space[2048];}@>
    // Insert read and close syscalls
<@\textcolor{red}{+  ~read(fd, space, sizeof(space));}@>
<@\textcolor{red}{+  ~close(fd);}@>
    setuid(0);
    exit(0);}
\end{lstlisting}
\caption{The modified malicious code.}
\label{fig:fig2}
\end{figure}

In this modified code, we insert the openat, read, and close syscalls related to a dummy file without affecting the original malware functionality. When executed, this modified code will generate a system call sequence consistent with our perturbation sequence. This demonstrates how our practical approach can guide the modification of the original code to match the desired perturbation sequence while maintaining the functionality of the program.

\section{System Design}\label{s:design}

% To ensure the practicality of generating adversarial samples, we only use insertion operations and ensure that the inserted behaviors also satisfy the order dependence relationship to generate more realistic adversarial samples. 
This section describes the overall architecture of the proposed sequence perturbation approach.
\begin{figure*}[h]
\centering
\includegraphics[scale=0.3]{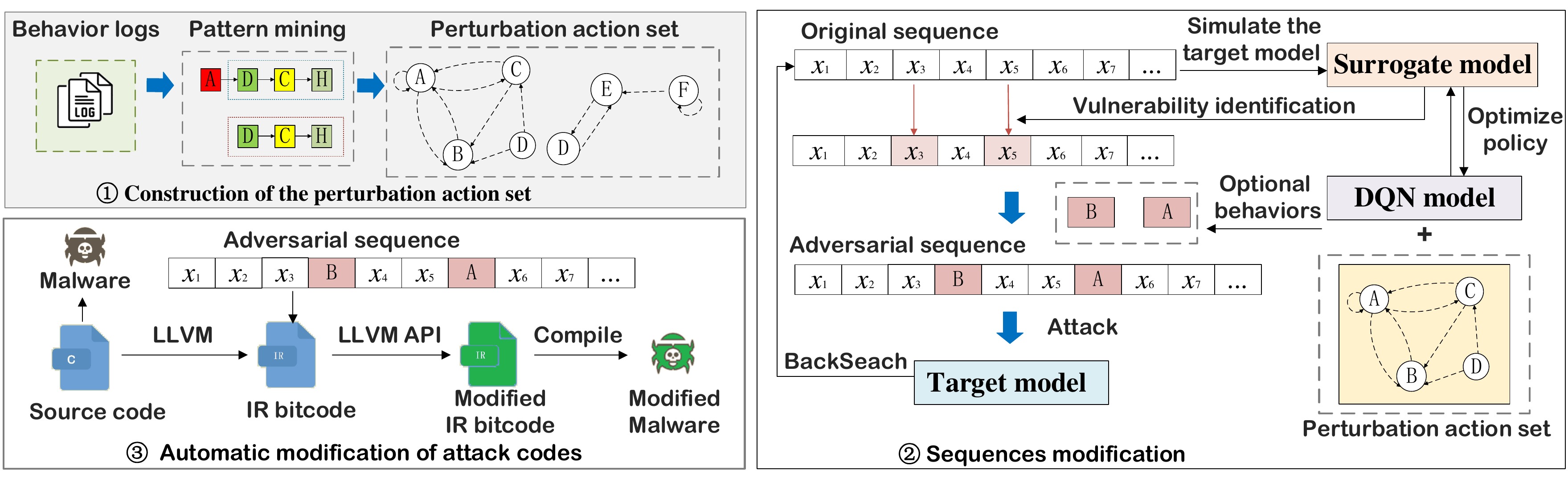}
\caption{\centering{The framework of our approach.}}
\label{fig_fram}
\end{figure*}

\subsection{Overview of the Proposed Approach}

We outline our proposed approach as illustrated in fig.\ref{fig_fram}, which consists of three steps: (1) construction of the perturbation action set; (2) adversarial sequences modification via backtracking search; and (3) automatic modification of attack code for adversarial sample generation.

%大写

\textbf{1) Construction of the perturbation action set}.
% To avoid affecting the functionality of programs, we choose not to replace or delete behavioral events. Instead, we aim to insert benign sequence fragments into the sequence. 
% In this step, frequent patterns are extracted from benign sequences using sequence pattern mining algorithms. Each extracted frequent pattern is then transformed into a benign behavior template, which can generate a specific sequence of behavioral events. By utilizing these behavior templates, diverse benign sequence fragments can be generated while ensuring their diversity. The generated sequence fragments are more representative of benign behavior due to the use of frequent patterns. These fragments can be utilized in subsequent adversarial sample generation, thus enhancing their effectiveness.
To avoid affecting the functionality of programs, we choose not to replace or delete behaviors. Instead, we aim to insert benign sequence fragments into the sequence. 
In this step, representative patterns are extracted from benign sequences using sequence pattern mining algorithms. The extracted patterns are then constructed as a perturbation action set, represented as directed graphs.
This set implies the range of possible perturbations that can be introduced by selectively inserting various benign behavior fragments into the original malicious behavior sequence. 
%we can then traversing directed graphs to generate benign behavior fragments by  the graphs 

%These directed graphs can be used 
%through various methods, such as depth-first search (DFS) or random selection of successor nodes, 
%to create diverse and logically related benign behavior sequence fragments. 

% \textbf{2) DQN-based approach for sequence perturbation}

% Due to the black-box nature of the attack, a surrogate detection model is first used to learn the behavior patterns of the sequence. When new behavior sequence is input, the surrogate detection model identifies potential anomalous situations. To perform adversarial modifications on the identified anomalous behavior, a gradient-based method is used to identify the optimal modification position. Then, a DQN (deep Q network) model learns to select the optimal adversarial action based on the current behavioral sequence and insertion position. The DQN continually improves the decision-making process by inputting the modified sequence to the surrogate detection model and receiving feedback. The specific perturbation process will be discussed in detail in Section 4.3.

\textbf{2) Sequences modification via backtracking search}.
Due to the nature of the black-box attack, a surrogate detection model is used to estimate the target model and detect malicious sequences. 
To perform adversarial modifications on the detected sequence, a gradient-based method is used to identify the optimal modification position. Additionally, a DQN model is trained to select the optimal behavior based on the current behavioral sequence. The DQN model continually improves the decision-making process by inputting the modified sequence into the surrogate detection model and receiving feedback.
When a new malicious sequence is input, the trained DQN model can iteratively modify the sequence in multiple steps. However, as mentioned in the introduction, inserting benign behaviors into a malicious sequence without proper constraints may result in semantically incorrect and invalid sequences. Therefore, we combine DQN with the perturbation action set and employ a heuristic, backtrack-based search strategy to generate practical behavior sequences. During the backtracking search process, we assign priorities to each possible behavior using the previously trained DQN model and adhere to the logical relationship constraints to generate feasible behavior sequences.

% To ensure the completeness of program functionality, we restrict the perturbation process to only insertion operations on the sequence. To improve the success rate of perturbation, the insertion operation may be multi-step. However, using a trained DQN model to find the best insertion behavior may not satisfy the dependency relationship among them. Therefore, we combine DQN with BGT and employ a heuristic, backtrack-based search strategy to generate a more compliant behavior sequence. Our approach aims to find a perturbation sequence that satisfies the attack success condition by inserting multiple behaviors into the optimal position in the sequence. At each recursive step, we assign priorities to each possible behavior using the previously trained DQN model and adhere to the automaton constraints to generate feasible behavior sequences. This method enhances the success rate of adversarial attacks while minimizing the perturbation and ensures compliance with the insertion behavior dependency relationship.

\textbf{3) Automatic modification of attack codes}.  
Our approach indirectly generates adversarial sequences by modifying malicious software. To verify the generation of attacks in real-world scenarios, a transformation approach based on LLVM is employed to map the sequence modifications back to the source code. Firstly, the source code is compiled into LLVM IR bitcode with debug information to identify the statements triggering the behaviors (i.e., system calls and API calls). Next, the program behaviors are tracked to determine the corresponding code locations. Then, the identified statements in LLVM IR bitcode are modified using the LLVM API. Finally, the modified LLVM IR is recompiled into an executable file to generate a new system call sequence that reflects the alterations made.

\subsection{Construction of the Perturbation Action Set}

%人工规则
%频繁模式，共现概率  注意大小写，时态
%相同的参数对象。
% 考虑 启发式算法，让他走完。
% To avoid impacting the functionality of programs, it is common practice to insert only benign sequence fragments into original malicious sequences. Furthermore, 
Inserting highly representative benign behavior fragments may further enhance the effectiveness of adversarial samples. To this end, we first collect a substantial amount of benign sequence data and perform pattern mining on it to discover subsequences exhibiting typical benign characteristics. Specifically, we utilize the frequent patterns mining approach to construct a perturbation action set:

%\textcolor{red}{Attackers can attempt to collect a substantial amount of benign sequences data and perform pattern mining on it to discover subsequences exhibiting typical benign characteristics.} 
%To this end, we utilize three sequence pattern mining approaches from benign samples to generate such benign sequence fragments and construct the perturbation action set: 

% This can be achieved by leveraging sequence pattern mining. we first propose three distinct approaches of sequence pattern mining , which can be utilized for generating adversarial samples.
% we first propose three distinct approaches for identifying critical subsequence behavior patterns within benign sequence samples, which can be utilized for generating adversarial samples. 

% \textbf{1) Mining frequent patterns}. 
By leveraging frequent pattern mining algorithms \cite{hegland2007apriori,zhang2008research}, 
% such as the Apriori algorithm \cite{hegland2007apriori} or the FP-Growth algorithm \cite{zhang2008research}, 
common subsequence patterns that occur frequently in benign sequence samples can be discovered. Mathematically, given a set of sequences $S = \{S_1, S_2, ..., S_n\}$ and a minimum support threshold $\delta$, the object is to identify all subsequences $S' = \{S'_1, S'_2, ..., S'_m\}$ such that $support(S'_i) \geq \delta$ for $i=1,...,m$, where $support(S'_i)$ represents the frequency of subsequence $S'_i$ occurring in the set of sequences $S$. The frequent subsequences obtained through this approach will better reflect the behavioral patterns of benign sequences, thus making them more representative.

These benign behavior patterns constitute a perturbation action set, we then use directed graphs $G$ to represent the sequential relationship of each pattern.

Specifically, for a pattern $L_i$ containing $m$ different behaviors, denoted as $B_i = \{b_{i1}, b_{i2}, \ldots, b_{im}\}$, we define the nodes $N_i = B_i$ , and the edges $E_i = \{(e_{ik}, e_{ik+1}) | 1 \leq k < m\}$ and construct the adjacency matrix $E_i$ for the graph $G_i$ as follows:

% Given a set of behavior patterns $\mathcal{P} = \{P_1, P_2, \ldots, P_n\}$ mined from benign behavior samples, we construct a directed graph, called a Behavior Generation Template (BGT), for each pattern $P_i = \{b_{i1}, b_{i2}, \ldots, b_{im}\}$. For each BST, we define the vertices $V_i = B_i = \{b_{i1}, b_{i2}, \ldots, b_{im}\}$, and the edges $E_i = {(b_{ik}, b_{ik+1}) | 1 \leq k < m}$, based on the sequence of behaviors in pattern $P_i$. Then, we construct the adjacency matrix $A_i$ for the graph $G_i$ as:

\begin{equation}
    E_i=\begin{pmatrix}e_{11}&e_{12}&\cdots&e_{1m}\\ e_{21}&e_{22}&\cdots&e_{2m}\\ \vdots&\vdots&\ddots&\vdots\\ e_{m1}&e_{m2}&\cdots&e_{mm}\end{pmatrix}
\end{equation}
Where $e_{xy} = 1$, if $(b_x, b_y) \in E_i$, representing that $b_y$ can appear after $b_x$; and $e_{xy} = 0$ otherwise.
% Each BGT is represented as a directed graph $G_i$ and its corresponding adjacency matrix $C_i$, where $c_{xy} = 1$ if $(b_x, b_y) \in E_i$ and $c_{xy} = 0$ otherwise. 

% By constructing directed graphs for each behavior pattern, we can traverse the graphs starting from different initial nodes and selecting successor nodes based on the graph structure. This enables us to generate various sequences of benign behaviors.

Through directed graph traversal starting from different initial nodes and selecting successor nodes based on the graph structure, diverse sequences of behaviors can be generated. It can introduce a wider spectrum of sequence variations while meeting the sequential relationship between benign behaviors. Inserting these benign sequence fragments into a malicious sequence contributes to the heightened complexity and semantic validity of adversarial samples.

For instance, consider the following directed graph, containing five behaviors: b1, b2, b3, b4, and b5. It has six behavioral sequential relationships:
b1→b2,
b1→b3,
b2→b4,
b2→b5,
b3→b4,
b4→b5.
By traversing this directed graph, we can generate the following possible behavior sequences:
Starting from b1, we can traverse the graph and get the sequences of (b1, b2), (b1, b2, b4), (b1, b2, b4, b5), (b1, b3), (b1, b3, b4), and (b1, b3, b4, b5). Similarly, starting from b2, we can traverse the graph and get the sequences of (b2, b4) and (b2, b4, b5). It demonstrates the flexibility and capability of directed graphs to represent behavior patterns and generate diverse benign behavior sequences.

% It demonstrates the flexibility and capability of directed graphs to capture complex behavior patterns and traverse them effectively to generate the required benign behavior sequences.

\subsection{DQN-based Approach for Sequence Perturbation}\label{subsec4.3}

%  图片和叙述不匹配
In this section, we introduce a DQN-based approach for sequence perturbation. The proposed approach aims to learn abnormal behavior characteristics of the sequences, identify vulnerable positions in the sequences for the insertion of perturbations and determine the modified strategy, ultimately generating adversarial samples. 
%Figure 4 illustrates the architecture of the proposed approach.

To perform a black-box attack, the target detection model is first queried by inputting the behavior sequences and observing the corresponding outputs. The surrogate model is then trained based on the input sequence and the predicted value of the target model, enabling it to successfully simulate the target model. Once a new behavior sequence is input to the surrogate model and if it detects anomalies, the sequence can be modified.

Next, the perturbation weight of different positions in the sequence is calculated to identify the most vulnerable modification positions. This is achieved by computing the gradient of the loss function for the surrogate model with respect to various positions of the behaviors in the sequence. A larger gradient value signifies a more significant impact of the input position on the loss function, which indicates a more susceptible modification position. 
% By identifying these vulnerable positions, we can effectively target and modify the behavior sequence to achieve the desired perturbation. The ultimate goal is to identify the optimal behavior to insert at the vulnerable positions to achieve the most effective perturbation. This process is achieved through utilizing a Deep Q-Network (DQN) model, a reinforcement learning approach that enables the attack process to adapt to the current behavior sequences.
% Our attack approach is focused on executing carefully chosen actions on the current behavior sequences based on optimal positions. The primary objective is to find a modification sequence that minimizes alterations to the original sequences, as opposed to making random modifications. Utilizing the DQN algorithm, an iterative learning method, can optimize the attack process, proving more effective than supervised or unsupervised learning approaches.
% The objective is to manipulate the original behavior sequence $S$ to influence the output of a target classifier $f(S)$, achieved by generating a new sequence $\hat{S}$ through modifications to $S$ such that $f(\hat{S}) \neq f(S)$. By applying the DQN algorithm to identify the optimal modification sequence, we can alter the classification without making excessive modifications to the original sequence. 

% After identifying vulnerable positions in the malicious sequence, the Deep Q-Network (DQN), a deep reinforcement learning technique is used to modify it to achieve the desired perturbation. 
After identifying vulnerable positions in the malicious sequence, the DQN model is employed to guide the modification process. The goal is to iteratively modify the malicious sequence until it is classified as benign, effectively altering the classification without introducing excessive modifications to the original sequence.
%aiming to achieve the desired perturbation while maintaining the malicious intent.
%Our focus is on executing carefully chosen actions on the sequence to minimize alterations and find an optimal modification sequence. This iterative learning approach is more effective than supervised or unsupervised learning \cite{}. 
% By iteratively modifying the malicious sequence until it is classified as benign, we can alter the classification without excessive modifications to the original sequence.

% Our approach is to insert a series of actions into the original behavior sequence to modify it until the classifier classifies it as benign. 
%The current state of the sequence depends on the previous states of all sequences and the modification operations performed on this sequence, which can be regarded as a large but finite Markov Decision Process (MDP). 
Specifically, the DQN model employs reinforcement learning to optimize the modification effectiveness. The modification process at each step can be stated as $s_{t+1} = M(s_{t}, a_{t})$, where $M$ is a function that applies the modification action $a_{t}$ to the current modification state $s_{t}$. Then, 
the Bellman equation is used to evaluate the Q-value \cite{mnih2013playing} of taking a specific action in a particular state:
%the  optimal action-value function $Q(s,a)$ 
\begin{equation}
Q(s,a) = [r + \gamma \mathop {\max }\limits_{a_{t+1}} Q(s_{t+1},a_{t+1})\mid s_t,a_t]
\end{equation}
where 
%$s$ represents the state, $a$ represents the action, 
$r$ represents the immediate reward obtained after taking action $a_t$ in state $s_t$, 
$s_{t+1}$ represents the next state after taking action $a$ 
%in the current state $s$
and $\gamma$ represents the discount factor.

To select the most effective attack action at each step, the DQN model uses the Q-value, which represents the expected long-term reward for a specific action in the current state.
During the training process, DQN optimizes the modification policy by observing the detection results of the surrogate model on the modified sequences, and receiving rewards. Next, we will discuss each component of our DQN-based approach.
%employs reinforcement learning to learn the attack process. This iterative learning algorithm 

\begin{figure}[h]
\centering
\includegraphics[scale=0.32]{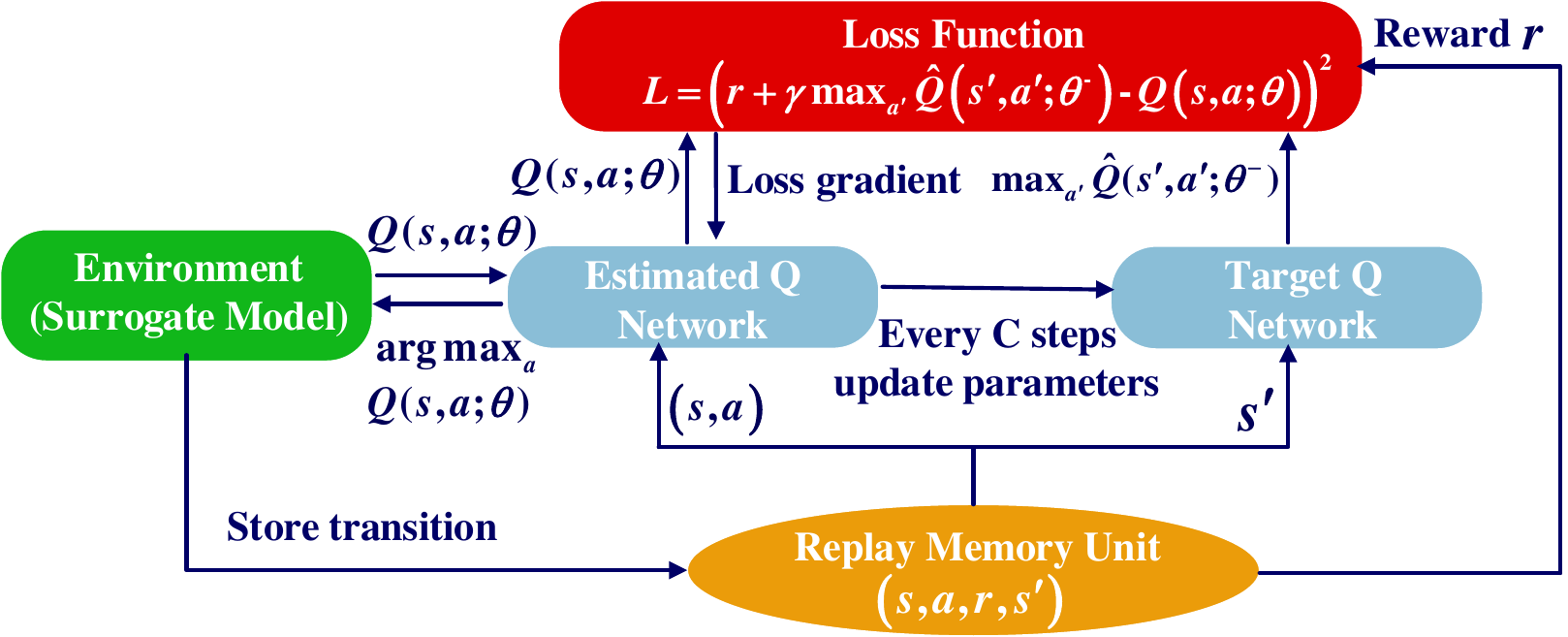}
\caption{\centering{Framework of DQN training algorithm.}}
\label{fig_dqnlearn}
\end{figure}

1) State space

%The perturbation learner modifies the sequences in a state space. 
The state space is represented by the behavioral sequences, denoted as $S$. Each sequence is encoded using one-hot encoding, where each value corresponds to a specific behavior.

Specifically, a state $s \in S$ is a behavior sequence of length $n$. Each element of the state vector $s$ corresponds to a specific behavior, represented as a number $s_i$
and $s_i \in \mathbb{R}$, for $1 \leq i \leq n$.  
The complete state space can then be represented as a set of all such vectors:
\begin{equation}
% S=\{s\in\mathbb{R}^n\mid1\leq i\leq n\}
S = \{s \mid s = (s_1, s_2, \ldots, s_n),\ s_i \in \mathbb{R},\ 1 \leq i \leq n\}
\end{equation}
The state space is processed by both the surrogate model and the DQN model. The surrogate model is responsible for detecting the malicious sequences, while the DQN model aims to modify the malicious sequences to mislead the surrogate model.

2) Action space

The action space is designed to represent the possible modification operations that can be applied to the behavior sequences. The action space is crucial in determining how effectively the modifications can change the classification results.

In our specific implementation, to maintain the original functionality of the program, we restrict the perturbation process to only inserting behaviors into the malicious sequences.
%The action space is discrete and consists of multiple actions, 
The action space, denoted as $A = \{a_1, a_2, \dots, a_n\}$, represents different insertion operations with corresponding behaviors $B = \{b_1, b_2, \dots, b_n\}$ at specific positions in the malicious sequences.
For each action $a_i \in A$, it can be represented as follows:
\begin{equation}
    \operatorname{Insert}(s,b_i,p)\to s'=[s_1,\ldots,s_{p-1},b_i,s_p,\ldots,s_k]
\end{equation}
where $s$ is the original malicious sequence with $k$ behaviors, $s'$ is the modified behavior sequence, $b_i$ is the inserted behavior, and $p$ is the position at which the behavior is inserted. The modified sequence $s'$ now contains $k+1$ behaviors, with the $b_i$ inserted at position $p$.

3) Reward function

The reward function is employed to guide the DQN model towards making effective modifications to the original malicious sequences that influence their classification results.
%We utilize the agent's loss function as the reward function and scale it by a specific factor.

Let us denote the prediction function of the surrogate model as $C(s; \phi)$, where $s$ represents the current state, and $\phi$ corresponds to the parameters of the surrogate model. Since the surrogate model is a binary classifier, the loss function $L_C(s, y; \phi)$ can be expressed using the categorical cross-entropy loss:
\begin{equation}
    L_C(s,y;\phi)=-\sum\limits_{i=1}^2y_i\log C(s;\phi)_i
\end{equation}
Here, $y$ is the true label of the behavior sequence, and $C(s; \phi)_i$ represents the predicted probability of the $i$-th class. 

The reward function $R(s, a, s'; \phi)$ can be formally defined as follows:

\begin{equation}
   R(s, a, s'; \phi) = \alpha \cdot (1- 2 \cdot L_C(s', y=0; \phi) )
\end{equation}

In this expression, $\alpha$ is a scaling factor that modulates the impact of the loss on the reward. By utilizing this reward function, the DQN model can be guided to minimize the loss of the surrogate model for the normal class (label 0), thereby effectively modifying malicious sequences to make them closer to benign sequences.

% The selection of the scaling factor $\alpha$ can be determined through meticulous experimentation or by considering the specific requirements of the log modification task at hand. By adjusting the value of $\alpha$, we can effectively calibrate the influence of the classifier's loss on the reward signal, thus ultimately steering the agent's learning process towards an optimal strategy.

4) Optimal position with gradient search

Instead of randomly modifying the behavior sequences, we can utilize the gradient of the surrogate model's loss function $L$ with respect to the positions in the sequence to find the most effective modification positions. Positions with higher gradient values have more significant influences on the loss function, indicating their vulnerability to modification. This approach ensures that the modifications made to the behavior sequences have the maximum impact on their classification results. Specifically, for a given behavior sequence
%behavior sequence $B = \{b_1, b_2, \dots, b_n\}$, the corresponding state 
%$s$ can be represented as a 
$s =\{s_1, s_2, \dots, s_n\}$. $\nabla L(s)$ can be the gradient vector of the loss function:
% let $\nabla L(B)$ be the gradient vector of the loss function for a given behavior sequence $B = \{b_1, b_2, \dots, b_n\}$, it can be represented as:
\begin{equation}
    \nabla L(s)=\left(\dfrac{\partial L}{\partial s_1},\dfrac{\partial L}{\partial s_2},\ldots,\dfrac{\partial L}{\partial s_n}\right)
\end{equation}

Next, we identify the element $i^*$ with the largest gradient magnitude. This index represents the position in the sequence that, when modified, will have the most significant impact on the classification result. It is represented as:
\begin{equation}
    i^*=\arg\max\limits_i|\nabla L(s)_i|,\quad1\leq i\leq n
\end{equation}

5) DQN agent and train strategy

To effectively modify the sequence, we adopt the DQN algorithm (Algorithm 1). This algorithm learns a policy to select the optimal insertion behavior based on the current sequence. 
%the modification to the sequence, based on the current state, which is the current sequence. 
Fig. \ref{fig_dqnlearn} illustrates the framework of DQN training algorithm.

The algorithm utilizes a deep neural network (estimated Q-Network) as a DQN agent to estimate the Q-function (line 4). 
This Q-function outputs the Q-values for different actions, guiding the agent to choose the action that maximizes the expected reward. 

The use of a neural network allows us to handle high-dimensional state spaces and complex mappings between states and actions, which are common in sequence modification tasks. It also enables the agent to generalize from seen states to unseen ones, an important capability given the vastness of possible sequences.

In each episode, the agent selects an action based on the $\varepsilon$-greedy policy, executes it, observes the reward and next state (lines 11-19). The agent then learns to maximize the expected cumulative reward, guiding its action selection towards sequence modifications that yield the highest rewards (lines 22-27). Our approach also employs two key mechanisms of DQN: experience replay and a target network. Experience replay (lines 20, 21) allows learning from past experiences, enhancing stability by breaking correlations in the data. The target network (line 28), providing a stable goal for the Q-function updates, further boosts stability. Each sequence is iteratively modified until it is recognized as benign or the maximum number of modification attempts is reached (lines 9, 30).

Through iterative learning, the agent develops a policy to make optimal sequence modifications. Thus, the DQN algorithm serves as an effective tool for sequence modification tasks.

\begin{algorithm}
\caption{ DQN training for sequence perturbation}
\begin{algorithmic}[1]
\State \textbf{Input}    Behavior sequences $S$, Modification times $T$
\State                  ~~~~~~~~~  surrogate model $F$, replay memory $D$
\State                  ~~~~~~~~~  Esitimated Q-Network $Q$, Target network $\hat{Q}$ 
\State Initialize $D$, $Q(\theta)$, $\hat{Q(\theta^{-})}$, where $\theta$, $\theta^{-}$ are weights
\State Perform one-hot encoding on the behavior sequences.
\For{episode $e$ $=1, M$}
    \For {$k$ $=1, count(S)$}
    \State state $s=S(k)$ 
    \For{time step $t=1, T$}
        \State Identify the vulnerable position $p$ in $s$.
        \State Probability $p$= random(0,1)
        \If{$p<\epsilon$}
        \State Randomly choose a action $a$
        \Else
        \State $a = \arg\max_{a'} Q(s, a'; \theta)$
        \EndIf
        \State Perform $a\rightarrow$ Insert$(s,b,p)$, 
        \State Get reward $r$ according to $F$
        \State Get the modified sequence (next state $s'$) 
        \State Add $(s, a, r, s')$ to $D$
        \State Randomly choose a subset $D'$ from $D$
        \If{the $e$ ends at step $j+1$}
        \State   $y_j=r_j$
        \Else
        \State   $y_j=r_j + \gamma \max_{a'} \hat{Q}(s'_j, a'; \theta^{-})$
        \EndIf
        \State Descend the gradient: 
        \Statex \;\;\;\;\;\;\;\;\;\;\;\;\;\;\;  $(y_j - Q(s_j, a_j; \theta))^2$ 
        \State Every $C$ steps, update $\hat{Q}$: $\theta^{-} \leftarrow \theta$
        \State $s \leftarrow s'$
        \If{$F(s)$ is normal}
        \State Break
        \EndIf
    \EndFor
    \EndFor
\EndFor
\end{algorithmic}
\end{algorithm}

\subsection{Backtracking Search Attack}\label{subsec4.4}

\subsubsection{Challenge}

After training the DQN model to learn the optimal policy for inserting behaviors, the ability of the DQN model to predict the optimal behavior insertions has been enhanced. The action-value function $Q(s,a;\theta)$, approximated by the DQN model, becomes more accurate and reliable in approximating the true optimal action-value function $Q^*(s,a)$.

\begin{equation}
    Q(s,a;\theta)\approx Q^*(s,a)\quad\text{}
\end{equation}

We use the predictions of the DQN model to prioritize available behaviors. By selecting the action with the highest predicted value at each step, we ensure that behaviors with the highest potential impact on the classifier's decision are considered first:

\begin{equation}
    \pi^*(s)=\arg\max_a Q^*(s,a)
\end{equation}

However, using the DQN model alone to select the best behaviors to insert in the behavior sequence may not account for the dependencies among the inserted behaviors. To illustrate this, let's consider the following scenario:

Suppose we have a set of behaviors $B = \{b_1, b_2, b_3\}$ with their corresponding dependencies defined by the rule $R$. According to $R$, $b_1$ must be followed by $b_2$, and $b_2$ must be followed by $b_3$ to maintain the dependency constraints.

If we use the DQN model alone, we might choose to insert the behaviors independently based on their estimated action-value function $Q(s_t, a; \theta_t)$. This is equivalent to solving the following optimization problem for each behavior:

\begin{equation}
a^*_{i,j} = \arg\max_a Q(s_{i,j}, a; \theta_t),  \quad j = 1, 2, \ldots, n
\end{equation}

However, this independent selection of behaviors may lead to a sequence that violates the dependency constraints specified by rule $R$. For example, the DQN model might choose to insert $b_1$ and $b_3$ into the original sequence without inserting $b_2$, resulting in an infeasible sequence:

\begin{equation}
\text{infeasible sequence} = \{\dots, b_1, \dots, b_3, \dots\}
\end{equation}

\subsubsection{Heuristic Backtracking Search Strategy}

To overcome the challenge as described, we propose a heuristic backtracking search algorithm. The heuristic backtracking algorithm achieves the optimal solution by systematically trying and discarding non-optimal solutions. In our approach, we employ a continuous process of attempting to modify malicious sequences while adhering to all behavior dependency constraints. Based on the perturbation set represented by the directed graph, as we traverse various modification paths, if the resulting adversarial samples cannot successfully evade classifiers, we backtrack to earlier states and explore alternative options. Through this iterative trial-and-error approach, we can ultimately generate adversarial samples that comply with all behavior dependency constraints. 

%achieves the optimal solution by systematically trying and discarding non-optimal solutions. 
%In our context, it is used to find a behavior sequence that complies with all dependency constraints. 

%回溯算法是一种通过试错解决问题的通用算法。在我们的上下文中，它用于查找符合所有依赖约束的行为序

As described in Algorithm 2, we combine the DQN model with the perturbation action set represented as directed graphs, which helps generate practical adversarial behavior sequences. It leverages the DQN model to predict the optimal behaviors to be inserted in the sequence (line 16), while concurrently considering the constraints and dependencies among behaviors as represented by the directed graphs (line 14).

During each iteration of the primary loop (line 8), the algorithm sorts the weights of the positions inserted into the sequence (line 9). Subsequently, it determines the potential behaviors to insert, either by acquiring all nodes in the directed graph (lines 11-12), or by obtaining all successor nodes from the $current\_node$ according to the directed graph (line 14). This ensures that the selected behaviors conform to the dependencies prescribed by the directed graph.

Following the selection of the optimal behavior using the DQN model (line 17), the behavior, denoted as $opt\_behav$, is inserted into the sequence, referred to as $best\_seq$ (line 19). The classification of the new sequence, termed as $new\_pred$, is then predicted (line 20). If the new sequence is classified as benign, the algorithm concludes successfully (lines 21-22). However, if the sequence does not satisfy the benign classification, the search change the next directed graph (line 25), thus permitting the insertion of other behaviors into the sequence. This process embodies the backtracking nature of the algorithm. If a malicious sequence of behaviors fails to be modified to a benign classification, the algorithm reverts its decisions and explores alternative sequences.

By balancing the predictive power of DQN and the structured representation of behavior dependencies in the directed graph, our heuristic backtracking search strategy presents a practical approach for generating behavior sequences. This strategy integrates the potential of heuristic search to prioritize promising paths and the adaptability of backtracking to reverse ineffective decisions, aiming to discover a sequence of behaviors that fulfill the dependency constraints and attain the desired benign classification.

\begin{algorithm} [h]
\caption{Backtracking search algorithm}
\begin{algorithmic}[1]
\State \textbf{Input}    $seq, ip\_weights, digraphs, dqn,$ 
\State                  ~~~~~~~~~  $max\_step, mod\_limit$
\Function{search()}{}
\State \textbf{global} $best\_seq \gets seq.copy()$
\If{$mod\_limit > max\_step$}
    \State \Return False
\EndIf
\For{$steps \in \{1, \dots, mod\_limit\}$}
    \State $pos \gets ip\_weights[seq].argsort()$
    \If{$steps == 1$}
        \State $nodes \gets findAllnodes(digraph)$
        \State $poss\_behav \gets nodes$
    \Else
        \State $poss\_behav \gets $
        \Statex \;\;\;\;\;\;\;\;\;\;\;\;\;\;\;\;$Traversal(digraph, current\_node)$
    \EndIf
    \State $behav\_weight \gets dqn.predict((seq))$
    \State $opt\_behav \gets {\operatorname{argmax}}$ $behav\_weight[poss\_beha]$
    \State $current\_node \gets opt\_behav$
    \State $best\_seq \gets insert(seq, [opt\_behav], [pos(steps)])$
    \State $new\_pred \gets classifier.predict(best\_seq)$
    \If{$new\_pred == benign$}
        \State \Return True
    \EndIf
\EndFor
\State \Return {$search(next~diagraph)$}
\EndFunction
\end{algorithmic}
\end{algorithm}

% The adversarial sample generation algorithm iteratively processes each automaton $a$ from the set of automata $A$. For each automaton, it first modifies the original sequence $S$ to create $S_{mod}$. Next, the surrogate model $M$ is used to predict whether $S_{mod}$ raises an anomaly flag. If no anomaly flag is raised, the heuristic backtracking search function $search$ is called with the modified sequence $S_{mod}$ and the current automaton $a$. If the search function returns a non-empty adversarial sequence $S_{adv}$, the algorithm terminates and returns the generated sequence. If no suitable adversarial sequence is found after processing all automata, the algorithm returns an empty sequence.

% \begin{algorithm}
% \caption{Real-time evasion attack}
% \begin{algorithmic}[1]
% % \Require original sequence $S$, set of automata $A$, surrogate model $M$, search function $search$
% \For {each automaton $a$ in $A$}
% \State $S_{mod} \gets S$
% \State $flag \gets M.predict(S_{mod})$
% \If{not $flag$}
% \State $S_{adv} \gets search(S_{mod}, a)$
% \If{$S_{adv}$ is not empty}
% \State \textbf{return} $S_{adv}$
% \EndIf
% \EndIf
% \EndFor
% \State \textbf{return} empty sequence
% \end{algorithmic}
% \end{algorithm}

\subsection{Automatic Modification of Attack Codes}\label{subsec4.5}
Our approach indirectly generates adversarial sequences by modifying malicious software. To verify that attacks can be generated in real scenarios, we propose a transformation approach to map sequence modifications back to the source code. Specifically, we make modifications to the system call sequences.

When modifying malicious code to generate adversarial samples, direct modification of the source code can face several challenges. For instance, the malicious code may be written in different programming languages, requiring different modification methods and techniques for each language. In addition, malicious code is often highly complex, making it difficult to understand and modify the source code. To overcome these challenges, we adopted a modification method based on LLVM \cite{lattner2004llvm}. It is because LLVM can compile source code from various languages into a unified Intermediate Representation (IR) bitcode. Additionally, IR bitcode has a relatively simple structure, consisting of basic blocks and instructions, making it easy to understand and manipulate.

% In order to identify the source code location associated with each system call in a traced sequence, 
We first compile the program with debug information and generate an executable file. Next, we utilize the strace tool with the -k option to generate a system call sequence that includes function stack information. The function stack information contains file offset addresses, which can be mapped to positions in the source code using the addr2line tool. This process allows us to obtain precise location information of the source code for each syscall within the entire sequence.
Next, we compile the source code into LLVM IR bitcode with debug information (using the -g option). This ensures that the generated LLVM IR statements include debug information such as the corresponding source code line numbers and file names. Next, we traverse the IR statements and search for IR statements related to system calls. Lastly, upon identifying all related statements, 
We utilize the LLVM API to automatically insert a custom external function (referred to as "insertSyscall") before each identified IR statement. Since we have already identified the source code location associated with each system call in a traced sequence, our custom function can determine whether it is necessary to insert system calls based on the source code location corresponding to the subsequent IR statement. In addition, the implementation of our custom functions takes into account the following two considerations:

% we utilize LLVM API (such as the IRBuilder class) to automatically insert the statements within the IR bitcode to generate extra system calls. 

% To implement modifications to behavior sequences (e.g. system calls), it is necessary to first determine the specific location of the relevant IR bitcode that triggers the system calls.  
% First, we compile the source code into LLVM IR bitcode with debug information (using the -g option). This ensures that the generated LLVM IR statements include debug information such as the corresponding source code line numbers and file names. 
% Next, we traverse the IR statements and search for IR statements related to behaviors(i.e., system calls or API calls). Upon identifying all related statements, we insert a custom function before them using the LLVM API to log debugging information. This allows us to track the program while also identifying the key code locations involved in these system calls and find the statements in LLVM IR that needs to be modified. Lastly, we utilize LLVM API (such as the IRBuilder class) to insert the statements within the IR bitcode. When inserting these calls, We specifically consider two cases that require attention:

\textbf{1) Inserting statements with related arguments}. If multiple IR statements need to be inserted and they are related (e.g., sys\_open and sys\_close), the choice of arguments should also be related to ensure proper functionality. When designing related arguments for inserting system calls, the dependencies between the calls are determined, and the function signatures are analyzed to understand the required arguments and data types. Consistent arguments are created across the related calls, maintaining dependencies and ensuring the output from one call is utilized as input for another.

\textbf{2) Control execution frequency and condition}. In our custom function inserted into the IR bitcode, we have implemented control over the execution count and conditions of system calls. Firstly, we define a configuration file to store the execution count of system calls and the source code locations where the system calls need to be inserted. Additionally, the same line of code may trigger multiple system calls, for example, when it is within a loop structure, we want to insert the system call only during certain iterations of the loop. Therefore, the configuration file also includes the counting conditions for the source code to trigger system calls. Once the configuration information is defined, our custom function can examine the insertion conditions specified in the configuration file and perform the system call insertion when the conditions are met.

After applying the desired changes, we recompile the modified LLVM IR bitcode into an executable file to generate a new system call sequence that reflects the alterations made.

\section{Evaluation}\label{s:evaluation}

To validate our approach, we evaluate the performance of our approach by answering the following research questions:
\begin{itemize}
\item  \textbf{RQ1: Effectiveness analysis}. \textit{ How effective is our approach against the detection models?}
\item  \textbf{RQ2: Modification efficiency comparisons}. \textit{How is the modification efficiency of our approach?}
\item  \textbf{RQ3: Effectiveness against defense approaches}. \textit{How effective is our approach against defense approaches?}
\item  \textbf{RQ4: Practicality assessment of attacks}. \textit{What is the feasibility of our approach and its advantages over other practical methods?}
\end{itemize}

\subsection{Environment}

To conduct our experiments, we employed Keras version 2.7 to construct our model. The hardware configuration of our system includes an AMD Ryzen 7 5800 8-Core Processor 3.40 GHz, NVIDIA 3060 graphics card, and 32.0 GB of RAM. 

\subsection{Dataset and Sequence Generation}

Our approach is evaluated on two public datasets and a synthesized dataset:

1) The first dataset is the ADFA-LD \cite{creech2013generation} dataset, which is a comprehensive collection of system call traces designed for evaluating intrusion detection systems (IDS) in Linux environments. It contains both normal and malicious behaviors, simulating various attack scenarios. Researchers  can leverage this dataset for training and testing IDS models, enhancing cybersecurity measures and fostering a deeper understanding of intrusion detection techniques in real-world Linux-based systems. 

2) To validate the practicality of our approach, we adversarially modify malware on the Linux x86 platform. As the corresponding source code of the ADFA-LD dataset is not provided, we download usable POCs (Proofs of Concept) from Exploit-DB, compile them into executable files, and trace the generated system call sequences using the strace tool. We then incorporate the traced call sequence data into ADFA-LD as a synthesized dataset.

3) The final dataset is the AndroCT dataset \cite{li2021androct}, which was specifically curated for Android malware detection. This dataset comprises more than 35,974 Android applications collected over a decade (2010-2019). It includes a diverse range of both benign and malicious applications, and the data is represented as API call sequences.  

We employed a fixed window grouping method to generate behavior sequences from these datasets. In the labeling process, we designated malicious sequences as "positive," while those benign sequences were labeled as "negative."

\subsection{Evaluation Method}

We introduce several commonly used metrics in the context of adversarial attacks:

Success rate (SR): The proportion of successful adversarial attacks where the prediction of a classifier is changed from the malicious class to the benign class. Mathematically, SR is defined as:
\begin{equation}
SR = \frac{\text{Number of successful attacks}}{\text{Total number of attacks}}
\end{equation}

Perturbation rate (PR): The average ratio of the number of perturbed (inserted or modified) elements in the sequence to the total length of the original sequence. This metric indicates the degree of distortion introduced to the input sequence by the adversarial attack. PR can be computed as follows:
\begin{equation}
PR = \frac{1}{N}\sum_{i=1}^{N} \frac{\text{Number of perturbed elements  } }{\text{Length of the original sequence } i}
\end{equation}

where $N$ is the total number of attacks.
\subsection{Target Model}

% ~~~~\textbf{1) Min Du et al. (2017) \cite{du2017deeplog}:} 
We choose four sequence-based models:

\textbf{1) LSTM-based model \cite{du2017deeplog}:} 
The approach called Deeplog utilizes LSTM networks to learn behavior patterns from behavior sequences. It represents each behavior pattern using one-hot vectors as input.

% \textbf{2) Sasho Nedelkoski et al. (2020) \cite{nedelkoski2020self}:} 
\textbf{2) Transformer-based model \cite{nedelkoski2020self}:} 
In this approach, the Transformer encoder with the multi-head self-attention mechanism is employed to learn contextual information from behavior sequences. The behavior sequences are represented as log vector embeddings.

% \textbf{3) Siyang Lu et al. (2018) \cite{ludetecting}:} 
\textbf{3) CNN-based model \cite{ludetecting}:}
This approach applies Convolutional Neural Networks (CNN) to capture complex relationships within behavior sequences, enabling the detection of anomalies in the behavior logs.

% \textbf{4) Amir Farzad et al. (2020) \cite{farzad2020unsupervised}:} 
\textbf{4) Autoencoder-based model \cite{farzad2020unsupervised}:} 
The proposed model introduces an unsupervised approach for behavior log anomaly detection, employing deep Autoencoder networks to detect anomalies in the behavior sequences.

\subsection{Baselines}
\subsubsection{Sequence Perturbation Approaches}

~~~~~\textbf{1) JD Herath et al. (2021) \cite{herath2021real}:} The proposed method, known as Log Anomaly Mask (LAM), utilizes deep reinforcement learning to generate adversarial samples by modifying behavioral sequences to evade anomaly detection. The method involves actions such as the replacement and dropping of behaviors within the sequences. LAM utilizes a sliding window approach to observe the sequence and determine the next modification action based on all the behaviors within the window. Therefore, if the step size of the sliding window is 1, this method needs to evaluate and modify each behavior in the sequence.

\textbf{2) Siyang Lu et al. (2023) \cite{lu2023black}:} It proposes two methods to attack deep neural networks used for log anomaly detection: an attention-based attacker (AA) and a gradient-based attacker (GA). Both methods also generate adversarial samples by dropping and replacing behaviors in the sequence. 
Both methods do not require modifying each behavior in the sequence, thereby improving modification efficiency and exhibiting a high success rate. However, applying these two methods to modify malware could potentially damage its functionality.

\subsection{Results}

\subsubsection{RQ1: Effectiveness analysis.}
\begin{figure}
    \begin{minipage}[t]{\linewidth}
        \centering
        \includegraphics[width=\textwidth]{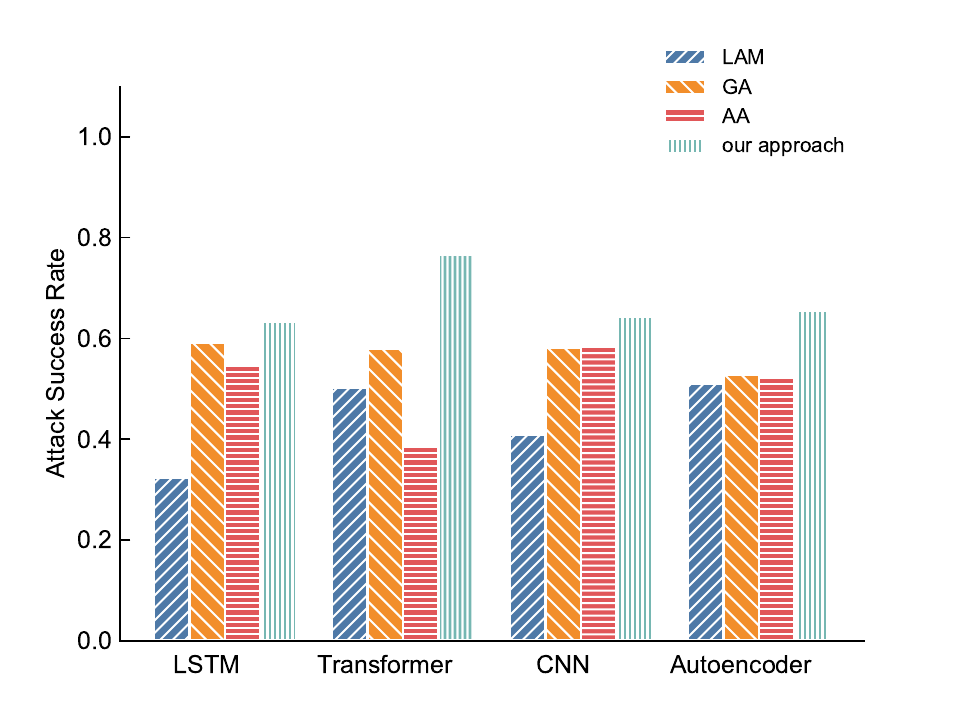}
        \centerline{(a) AndroCT  }
    \end{minipage}%
    
    \begin{minipage}[t]{\linewidth}
        \centering
        \includegraphics[width=\textwidth]{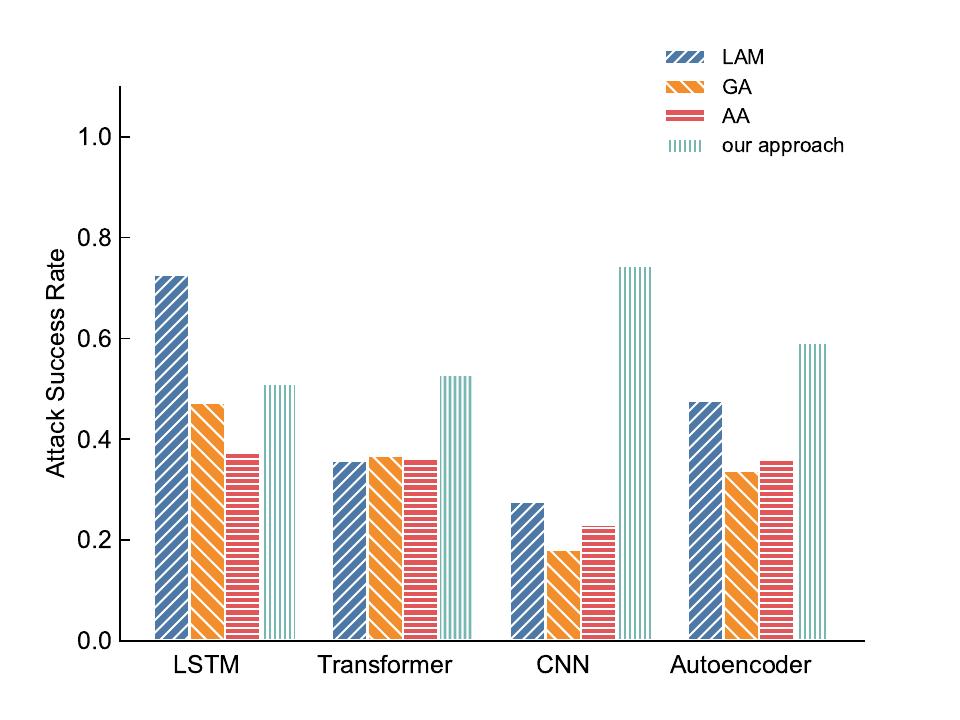}
        \centerline{(b) ADFA-LD}
    \end{minipage}
    \caption{Effectiveness analysis on two datasets.}
    \label{Effectiveness}
\end{figure}

% To evaluate the effectiveness of our proposed approach, 
We evaluate our approach on AndroCT and ADFA-LD datasets. We compared the performance of our approach with the state-of-the-art baseline methods in terms of success rate. The result is shown in Fig\ref{Effectiveness}. While our approach may not always achieve the highest success rate compared to the baselines \cite{herath2021real,lu2023black} that employ deletion and replacement operations, it still demonstrates competitive performance without disrupting the original functionality of the sequence and maintaining the integrity of the original sequence.

For instance, when evaluated on the AndroCT dataset, our approach achieves a success rate (SR) of 64.4\% against the CNN-based model 
% \cite{ludetecting}
, while the  baseline approach of GA 
% \cite{lu2023black} 
gets a SR of 58.1\%. Similarly, on the ADFA-LD dataset, the SR of our approach is 59.1\% against the Autoencoder-based model 
% \cite{farzad2020unsupervised}
, whereas the LAM approach 
% \cite{herath2021real} 
achieves a success rate of 47.6\%. These results indicate that our approach is effective in generating valid adversarial sequences that can evade detection by only inserting new behaviors, while preserving the integrity of the sequence. That is because our approach can dynamically select the most suitable perturbation behaviors and the optimal position to insert according to the current sequence. Additionally, the results demonstrated that the generated adversarial sequences exhibit a comparable degree of transferability across different classifiers and datasets, similar to the baseline approaches. This suggests that our approach is effective in a wide range of scenarios and applications, further emphasizing its effectiveness in real-world settings.

Furthermore, we analyze the legitimacy of the generated adversarial sequences compared to the baseline methods. Our approach effectively maintains the sequential order of the inserted behaviors in the adversarial sequence, resulting in more realistic adversarial sequences. This is particularly important for preserving the functionality of the program and reducing suspicion due to unnatural or inconsistent behavior patterns. The result revealed that the adversarial sequences generated by our approach were more realistic compared to those produced by the baseline methods.

For example, consider a malicious system call sequence: \texttt\{..., mknodat, execve, io\_cancel, setxattr, ...\} in the ADFA-LD dataset. Our approach inserts semantically valid benign system call sequences, such as \texttt\{getuid, lsetxattr\}, resulting in the adversarial sequence \texttt\{..., mknodat, execve, io\_cancel, getuid, lsetxattr, setxattr, ...\}. In contrast, the baseline approach of LAM replaces system calls, leading to a less practical sequence like \texttt\{..., mknodat, lsetxattr, lsetxattr, setxattr, ...\}.

The following table provides a comparison of the original sequence, our generated adversarial sequence, and sequences generated by two baseline methods, illustrating how our approach generates more practical adversarial sequences by inserting semantically valid benign sequences, as opposed to the less practical sequences generated by the baseline methods.

\begin{table}[h]
\renewcommand\arraystretch{1.2}
\centering
\caption{Comparison of original sequence, and adversarial sequence generated by different methods.}
\begin{tabular}{|l|l|}
\hline
\textbf{Method} & \textbf{System Call Sequence} \\
\hline

\multicolumn{1}{|l|}{\multirow{2}{*}{Original sequence}} & \texttt{...,mknodat,execve,io\_cancel} \\
\multicolumn{1}{|l|}{}   & \texttt{setxattr,...} \\

\hline
AA or GA (Delete) & \texttt{mknodat,io\_cancel,setxattr,...} \\
\hline
\multicolumn{1}{|l|}{\multirow{2}{*}{LAM (replace)}} & \texttt{...,mknodat,lsetxattr,lsetxattr} \\
\multicolumn{1}{|l|}{}   & \texttt{setxattr,...} \\
\hline
\multicolumn{1}{|l|}{\multirow{2}{*}{Our approach}} & \texttt{...,mknodat,execve,io\_cancel} \\
\multicolumn{1}{|l|}{}   & \texttt{getuid,lsetxattr,setxattr,...} \\
\hline
\end{tabular}
\end{table}

In conclusion, our approach generates highly effective adversarial sequences that preserve the logical relationships among behaviors while evading detection. Our approach generates more realistic and semantically coherent adversarial sequences compared to the baseline methods, which is particularly important for maintaining the original functionality and reducing suspicion.

% \begin{table*}[ht]
% \renewcommand\arraystretch{1.2}
% \centering
% \caption{The robustness to adversarial attacks of different methods.}
% \label{tab_rob}
% \begin{tabular}{cp{12em}ccccc}
%     \toprule
%     \multirow{2}[3]{*}{Datasets} & \multirow{2}[3]{*}{Taget model} & \multicolumn{5}{p{25em}}{~~~~~~~~~Perturbation rate of different methods} \\
% \cmidrule{3-7}    & \multicolumn{1}{c}{} & ~~~~LAM~~~~   &  ~~~~AA~~~~  & ~~~~GA~~~~ & random insert& our approach \\
%     \midrule
%     \multirow{3}[3]{*}{Dataset1} & LSTM-based Model            & 0.992 & 0.931 & 0.885 & 0.931 & 0.885  \\
%                                  & Transformer-based Model     & 0.982 & 0.993 & 0.991 & 0.931 & 0.885  \\
%                                  & CNN-based Model             & 0.982 & 0.993 & 0.991 & 0.931 & 0.885  \\
%                                  & Autoencoder-based Model     & 0.982 & 0.993 & 0.991 & 0.931 & 0.885  \\
%     \midrule
%     \multirow{3}[3]{*}{Dataset1} & LSTM-based Model            & 0.992 & 0.931 & 0.885 & 0.931 & 0.885  \\
%                                  & Transformer-based Model     & 0.982 & 0.993 & 0.991 & 0.931 & 0.885  \\
%                                  & CNN-based Model             & 0.982 & 0.993 & 0.991 & 0.931 & 0.885  \\
%                                  & Autoencoder-based Model     & 0.982 & 0.993 & 0.991 & 0.931 & 0.885  \\

%    \bottomrule
% \end{tabular}%
% \end{table*}

\subsubsection{RQ2: Modification efficiency comparisons}

% To evaluate the efficiency of our proposed approach in modifying behavior sequences, 
We compared the perturbation rate (PR) of our approach with the baselines on two datasets to analyze the efficiency of our approach, as is shown in TABLE \ref{tab_2}.

\begin{table}[ht]
\renewcommand\arraystretch{1.2}
\centering
\caption{Perturbation rate of different attack methods.}
\label{tab_2}
\begin{tabular}{ccc}
    \toprule
 \multirow{1}[1]{*}{Datasets} & \multirow{1}[1]{*}{Method} &  \multirow{1}[1]{*}{ Perturbation rate } \\
    \midrule
    \multirow{3}[5]{*}{AndroCT}  & LAM            & 20.3\% \\  %4.06
                                 & AA             & 23.4\% \\  %4.69
                                 & GA             & 22.9\% \\  %4.58 
                                 & Our approach   & 21.1\% \\  %4.21
    \midrule
    \multirow{3}[5]{*}{ADFA-LD}     & LAM            & 20.1\%\\   %4.01
                                 & AA             & 19.5\% \\  %3.89
                                 & GA             & 20.4\% \\  %4.08
                                 & Our approach   & 18.5\% \\  %3.69

   \bottomrule
\end{tabular}%
\end{table}

Our approach demonstrates a comparable PR compared to other baseline methods on both two datasets. For example, on the ADFA-LD dataset, our approach achieves a PR of 18.5\%, whereas the baseline method of GA achieved a PR of 20.4\%. This shows that our approach introduces fewer modifications to the original sequences while still effectively evading detection. In contrast, the PR of our approach is higher than the LAM on the AndroCT dataset, but the difference is not significant. Our approach has a PR of 21.1\%, while the LAM method has a PR of 20.3\%. Despite the higher PR, our approach maintains the functionality of the program and can be used to guide the modification of source code.

% Furthermore, our method was compared with a random insertion method, where normal behavior segments were inserted into the sequences at random positions. Our approach outperformed the random insertion method in terms of PR on both datasets. For example, on Dataset B with the target model of Sasho Nedelkoski et al. (2020) \cite{nedelkoski2020}, our method achieved a PR of WW\%, while the random insertion method had a PR of VV\%. This comparison demonstrates that the adversarial sequences generated by our method are not only more effective in evading detection but also with a lower PR, as opposed to random insertion methods.

These results highlight the efficiency and practicality of our approach in generating adversarial behavior sequences that can effectively evade detection with lower perturbations on both datasets.

\subsubsection{RQ3: Effectiveness against defense approaches}

% To evaluate the effectiveness of our proposed approach against various defense approaches, 

We also assess the performance of the adversarial sequences' evasive capabilities against various types of advanced defense methods. This analysis would provide valuable insights into the effectiveness and versatility of our approach. We conducted our experiments on several typical defense approaches against behavior log adversarial attacks. The details are as follows:  

\textbf{1) Sequence Squeezing \cite{rosenberg2021sequence}:} Sequence squeezing reduces the space of perturbations by grouping behaviors with similar semantic features to defend against adversarial attacks. For example, the syscalls sys$\_$writev and sys$\_$write can be merged into a group.

% Sequence squeezing reduces the search space available to an adversary by merging similar semantic features into a single representative feature. For instance, the syscalls “sys$\_$read” and “sys$\_$read” can be merged into a system operation behavior of “reading file”. To implement this approach, we use word2vec to represent the syscall names as word embeddings and cluster similar syscalls with the same semantics (by using Euclidean distance). Therefore, different merged groups can represent different operation behaviors.}

\textbf{2) Adversarial Learning \cite{szegedy2013intriguing}:} Adversarial learning incorporates adversarial examples into the training data, enabling the classifier to recognize and counter adversarial attack patterns.

% Adversarial learning adds adversarial samples to the training set, which can make the classifier learn the distribution of adversarial samples, thereby defending against adversarial attacks. We generate malicious adversarial samples according to Algorithm \ref{alg2} and add them to our training dataset. In addition, we label these samples abnormal. Then, we train the classifier using the dataset and analyze the detection performance. }

\textbf{3) Defense Sequence-GAN \cite{rosenberg2019defense}:} Defense Sequence-GAN achieves defense by training two GANs to approximate the samples to be detected to undisturbed sample distribution

% We employ SeqGAN to implement this approach. We train a benign SeqGAN and malicious SeqGAN using the unperturbed dataset and generate benign samples and malicious samples, respectively. When an input sequence emerges, we choose the generated unperturbed sequence nearest the perturbed sequence (calculated by Euclidean distance) and feed it to the classifier.}

%这种方法背后的原因依赖于这样一个事实：熟练的攻击者可以躲避检测器，我们的目标是通过添加检测器来增加发起对抗性攻击的难度。换句话说，我们希望只有当所有三个探测器都成功地被攻击欺骗并输出较低的分数时，系统才能被躲避。在现实世界中，很难欺骗所有探测器。由于三个检测器的输入不同，为了欺骗所有检测器，攻击者应该改变更多特征，这可能会破坏底层逻辑和攻击功能。

% These defense methods can make a classifier more robust to adversarial examples, without explicitly trying to detect them. In this experiment, we select LSTM-based model as target detection model and evaluate our approach on ADFA-LD dataset. 

These defense methods aim to enhance the robustness of a classifier against adversarial samples, without the explicit goal of detecting them. In our experimental setup, we select the LSTM-based model as the target detection model and conducted evaluations using the AndroCT dataset.

\begin{table}[ht]
\renewcommand\arraystretch{1.2}
\centering
\caption{The robustness to adversarial attacks of different methods.}
\label{tab_rob}
\begin{tabular}{cp{1em}cccc}
    \toprule
    \multirow{2}[3]{*}{Target defense technology} & \multicolumn{4}{p{16em}}{~~~~Success rate of different methods} \\
    \cmidrule{2-5}  & LAM & ~~~~GA & AA & our approach   \\
    \midrule                                                                       
                                  Adversarial Learning        & 29.6\% & ~~~~51.5\% & 48.7\% & 60.6\% \\
                                  Sequence Squeezing          & 30.3\% & ~~~~52.7\% & 49.2\% & 61.2\% \\
                                  Defense Sequence-GAN        & 31.2\% & ~~~~54.2\% & 50.8\% & 62.1\% \\

   \bottomrule
\end{tabular}%
\end{table}

% Overall, after implementing defensive measures, the success rates of all attack methods decreased. However, in the case of Adversarial Learning method, our approach achieves a success rate of 0.606. Given the reputation of adversarial learning as a strong defense measure, this result underscores our method's ability to elude even well-established defenses. This implies that, even though adversarial learning augments the training set's robustness by including adversarial samples, our approach remains effective.

% when evaluating against the Sequence Squeezing defense method, our approach still achieves a success rate of 0.612, slightly outperforming other techniques, namely LAM (0.303), GA (0.527), and AA (0.492). This suggests that while sequence squeezing consolidates similar semantic features to reduce the adversarial search space, our method can effectively bypass such a defense mechanism.

% Regarding the Defense Sequence-GAN method, our approach achieved a success rate of 0.621, the highest among the tested methods. This shows that, despite the Defense Sequence-GAN's efforts to emulate the distribution of unperturbed behavior sequences, our technique remains resilient.

The results are presented in TABLE 3. Overall, while the implementation of various defensive measures generally reduces the success rates of attack methods, our approach consistently demonstrates robustness across different defenses. Particularly noteworthy is its performance against the Adversarial Learning defense, achieving a success rate of 60.6\%, emphasizing its prowess in bypassing even well-regarded defensive measures. When pitched against the Sequence Squeezing defense, our approach not only outperforms other techniques such as LAM (30.3\%), GA (52.7\%), and AA (49.2\%) with a success rate of 61.2\%, but it also showcases its capacity to navigate around defenses that aim to shrink the adversarial search space by merging similar semantic features. Furthermore, against the Defense Sequence-GAN, our approach also stands out with the highest success rate of 62.1\%.

% Overall, these results illustrate the significant resilience and adaptability of our approach in the face of a range of defensive techniques.

The performance of our approach against various state-of-the-art defense techniques reaffirms its robustness and effectiveness. This is attributed to the fact that our approach effectively changes the behavior pattern of the original malicious sequence by inserting benign behavior segments. In addition, the perturbations introduced are minimal, ensuring the adversarial samples remain similar to the original sequences while achieving their adversarial attacks. 

% These findings provide a foundation for further research into refining adversarial techniques and improving current defense strategies.

% \begin{table}[ht]
% \renewcommand\arraystretch{1.2}
% \centering
% \caption{The robustness to adversarial attacks of different methods.}
% \label{tab_rob}
% \begin{tabular}{cp{1em}cccc}
%     \toprule
%     \multirow{2}[3]{*}{Target defense technology} & \multicolumn{4}{p{16em}}{~~~~Success rate of different methods} \\
%     \cmidrule{2-5}  & LAM & ~~~~GA & AA & our approach   \\
%     \midrule
%                                   Adversarial Learning        & 0.992 & ~~~~0.931 & 0.931 & 0.931 \\
%                                   Sequence Squeezing          & 0.982 & ~~~~0.993 & 0.931 & 0.931 \\
%                                   Defense Sequence-GAN        & 0.982 & ~~~~0.993 & 0.931 & 0.931 \\

%    \bottomrule
% \end{tabular}%
% \end{table}

\subsubsection{RQ4: Practicality assessment of attacks}
When assessing practicality, we mainly analyze the difficulty of implementing adversarial attacks. Our approach avoids direct tampering with logs, thereby reducing the complexity of attacks. Secondly, our approach involves adding benign behaviors to the generated sequences of malicious behaviors rather than performing deletion or replacement operations. It is relatively straightforward to implement and ensures the preservation of the original program's functionality. To further demonstrate the advantages and superiority of the proposed approach, we also compare it with other practical adversarial attack approaches (those that modify the malware without affecting its functionality) on the same datasets. The baselines are as follows: 

\textbf{1) Decision-Based Attack: \cite{rosenberg2020query}:} This method iteratively selects random insertion positions within the malicious sequence. Then, it inserts random behaviors or behaviors from the benign sequence into the malicious sequence.

\textbf{2) Feature Level Attack: \cite{fadadu2020evading}:} This method also iteratively modifies the sequence by randomly selecting a position within the sequence, but adds a behavior with the intention of maximizing a predefined classification score difference between benign and malicious samples.

% maximize the difference between the score function of benign samples and that of malicious samples, until the sequence is classified as normal. The motivation behind this is to introduce a characteristic that is more likely to appear in benign executable files than in malicious ones, allowing malicious sequences to mimic benign behavior.

\begin{figure}
    \begin{minipage}[t]{0.95\linewidth}
        \centering
        \includegraphics[width=\textwidth]{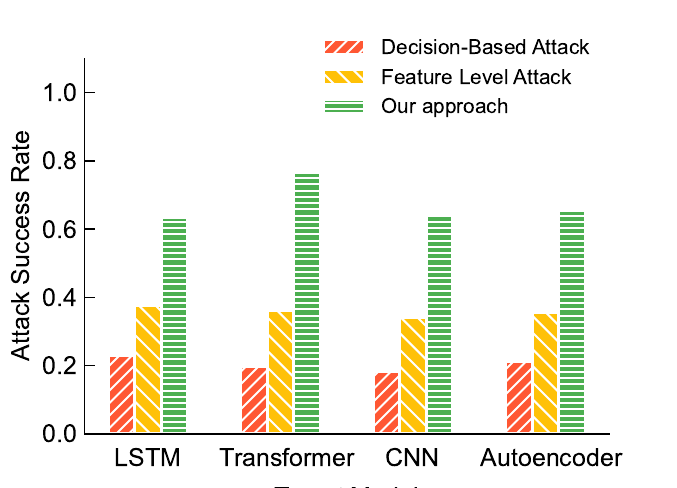}
        \centerline{(a) AndroCT  }
    \end{minipage}%
    
    \begin{minipage}[t]{0.95\linewidth}
        \centering
        \includegraphics[width=\textwidth]{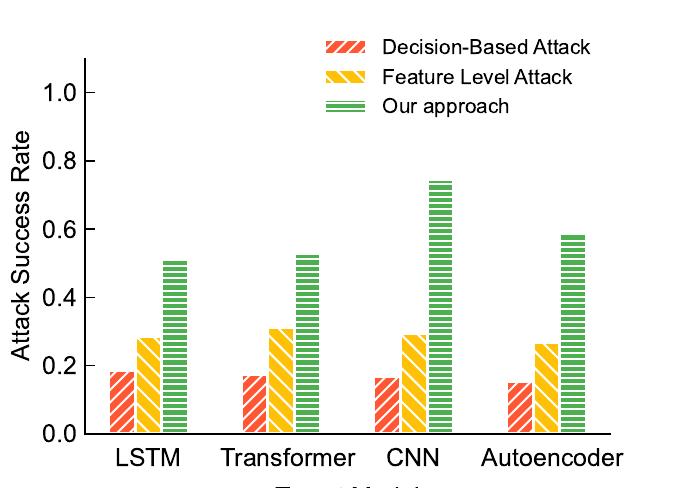}
        \centerline{(b) ADFA-LD}
    \end{minipage}
    \caption{Effectiveness analysis of practical approaches.}
    \label{practical}
\end{figure}

We compare our approach with the baseline methods in terms of attack success rate and sequence semantic legitimacy. The result is shown in Fig.\ref{practical}. Firstly, our approach achieves the highest success rate in attacking the four target models on two datasets, with average attack success rates of 67.5\% and 59.3\%, respectively. For the baselines, their performance in terms of attack success rate is significantly lower compared to our approach. For instance, the Decision-Based Attack method only achieves  the average attack success rate of 16.7\% on the ADFA-LD dataset. This can be attributed to the inherent randomness of their insertion strategies into the sequence. Consequently, these methods often fail to ascertain and capitalize on the optimal position or behavior for insertion, thereby not fully perturbing the decision-making process of the classifier. However, our approach can dynamically select the most suitable perturbation behaviors and the optimal position to insert according to the current sequence. 

In addition, similar to the sequence legitimacy analysis in section 4.6.1, these methods cannot always observe all behavioral dependency constraints like our approach during each modification of the sequence. For instance, [4] inserted an illegitimate benign fragment \{read, capset, open\} into the sequence. However, our approach applies the heuristic backtracking search algorithm, which searches based on semantic validity constraints during the modification process, and the final sequence is semantically valid.

\subsubsection{Case Study}

%在这一个章节，我们给出一个具体的修改例子，用于说明我们攻击的可行性。 从Exploit-Db上下载的poc文件22105.c，进行编译后。通过strace 跟踪得到的系统调用序列为。\{openat, mmap, getpid, clone, clock_nanosleep, restart_syscall  exit_group\}。通过我们方法，在该序列插入\{setxattr, setxattr, getuid, getuid\}。得到的对抗性序列为 \{setxattr, setxattr, openat, mmap, getuid, getuid, getpid  clone, clock_nanosleep, restart_syscall, exit_group \}。我们首先将程序源码编译为 IR bitcode。部分代码如图所示。然后，我们修改IR bitcode，在每个可能产生系统调用的statement之前插入了一个自定义的外部函数，部分代码如图所示。然后将系统调用的插入条件写入配置文件中用于外部函数的修改。我们将。最后we compile the modified LLVM IR bitcode into an executable file 重新用strace进行跟踪对修改后的程序进行验证。得到的跟踪系统调用序列如图所示。结果表明，我们方法在真实程序中的可用性。

% 在这个章节中，我们提供了一个具体的修改示例，以说明我们攻击的可行性。我们把从Exploit-Db上下载的poc文件,22105.c 进行了编译。通过使用"strace"工具来跟踪程序执行期间的系统调用序列，我们获得了以下序列：{openat, mmap, getpid, clone, clock_nanosleep, restart_syscall, exit_group}。为了实现这个修改，我们首先将程序源代码编译为LLVM IR（中间表示）的bitcode形式。以下是部分代码的示例。然后，我们修改了LLVM IR bitcode，在每个可能产生系统调用的statement之前插入了一个自定义的外部函数。以下是修改后的部分代码的示例。我们将系统调用的插入条件写入配置文件，用于外部函数对程序进行修改。最后，我们将修改后的LLVM IR bitcode编译为可执行文件，并使用"strace"对修改后的程序进行验证。验证结果显示，我们的方法在真实程序中是可行的。
We modified the source code of 80 usable POCs downloaded from Exploit-DB to generate corresponding adversarial sequences. Subsequently, all of these sequences were successfully compiled into executable files and can bypass the classifiers.
Taking the PoC file "22105.c" as an example, we first compiled it and traced the system call sequence during program execution using the strace tool. The observed sequence consists of the following system calls: \{openat, mmap, getpid, clone, clock\_nanosleep, restart\_syscall, and exit\_group\}.

% In this section, we aim to demonstrate the feasibility of our attack by providing a specific modification example. To achieve this, we compiled the Proof-of-Concept (PoC) file "22105.c" obtained from Exploit-Db and traced the system call sequence during program execution using the strace tool. The observed sequence consists of the following system calls: \{openat, mmap, getpid, clone, clock\_nanosleep, restart\_syscall, and exit\_group\}.

\begin{figure}[h]
\centering
\begin{lstlisting}[language={llvm},numbers=left, numberstyle=\tiny,keywordstyle=\color{blue!70},commentstyle=\color{red!50!green!50!blue!50},frame=single, rulesepcolor=\color{red!20!green!20!blue!20},escapeinside={<@}{@>}]
define dso_local i32 @main() #0 !dbg !69 {
  ...
  %9 = call i32 (ptr, i32, ...) 
  @open(ptr noundef 
  @.str, i32 noundef 0), !dbg !85 
  store i32 %9, ptr %4, align 4, !dbg !86
  %10 = load i32, ptr %4, align 4, !dbg !87
  %11 = call ptr @mmap(ptr noundef null, 
  i64 noundef 40960, i32 noundef 0, 
  i32 noundef 2, i32 noundef %10, 
  i64 noundef 0) #4, !dbg !88
  %12 = ptrtoint ptr %11 to i32, !dbg !89
  store i32 %12, ptr %2, align 4, !dbg !90
  %13 = call i32 @getpid() #4, !dbg !91
  store i32 %13, ptr %6, align 4, !dbg !92
  %14 = call i32 @fork() #4, !dbg !93
  %15 = icmp ne i32 %14, 0, !dbg !93
  br i1 %15, label %34, label %16, !dbg !95
  ...
\end{lstlisting}
\caption{ The malicious IR bitcode of "22105.c".}
\label{fig6}
\end{figure}

\begin{figure}[h]
\centering
\begin{lstlisting}[language={llvm},numbers=left, numberstyle=\tiny,keywordstyle=\color{blue!70},commentstyle=\color{red!50!green!50!blue!50},frame=single, rulesepcolor=\color{red!20!green!20!blue!20},escapeinside={<@}{@>}]
define dso_local i32 @main() #0 !dbg !69 {
  ...
   %9 = call i32 (ptr, i32, ...) 
   @open(ptr noundef 
   @.str, i32 noundef 0), !dbg !85
   store i32 %9, ptr %4, align 4, !dbg !86
   %10 = load i32, ptr %4, align 4, !dbg !87
<@\textcolor{red}{+ call void @insertSyscall(i64 1), !dbg !88}@>
   %11 = call ptr @mmap(ptr noundef null, 
   i64 noundef 40960, i32 noundef 0, 
   i32 noundef 2, i32 noundef %10, 
   i64 noundef 0) #4, !dbg !88
   %12 = ptrtoint ptr %11 to i32, !dbg !89
   store i32 %12, ptr %2, align 4, !dbg !90
<@\textcolor{red}{+ call void @insertSyscall(i64 2), !dbg !91}@>
   %13 = call i32 @getpid() #4, !dbg !91
   store i32 %13, ptr %6, align 4, !dbg !92
<@\textcolor{red}{+ call void @insertSyscall(i64 3), !dbg !93}@>
   %14 = call i32 @fork() #4, !dbg !93
   %15 = icmp ne i32 %14, 0, !dbg !93
   br i1 %15, label %34, label %16, !dbg !95
  ...
\end{lstlisting}
\caption{The modified malicious IR bitcode of "22105.c".}
\label{fig7}
\end{figure}

Next, using our approach, we insert the system calls \{setxattr, setxattr, getuid, getuid\} into the original sequence. The resulting adversarial sequence becomes: \{setxattr, setxattr, openat, mmap, getuid, getuid, getpid, clone, clock\_nanosleep, restart\_syscall, exit\_group\}.

To implement this modification, the program source code is first compiled into LLVM IR bitcode. The partial code is shown in Fig.\ref{fig6}. Then, the LLVM IR bitcode is modified by automatically inserting custom external functions before each statement that may generate expected system call, as shown in Fig.\ref{fig7}.  Finally, the modified LLVM IR bitcode is recompiled into an executable file and generates the expected system call sequence. The result underscores the feasibility of our approach in manipulating the system call behavior of a program.

\section{Related Work}\label{s:related-work}
 We provide an overview of the related work in two primary domains: deep learning-based malware detection and adversarial attacks on deep learning models.

\subsection{Deep Learning-based Malware Detection}

% Malware detection is a crucial task in safeguarding computer systems \cite{aboaoja2022malware}. In this field, static analysis \cite{omar2022static} and dynamic analysis \cite{li2022novel,mahdavifar2022effective} are two commonly employed methods. Static analysis involves examining malicious software samples through code analysis and feature extraction. However, it can be hindered by techniques like code obfuscation and encryption \cite{or2019dynamic}, making it challenging to uncover the true intent of the malware. In contrast, dynamic analysis entails executing malware samples in a controlled environment while monitoring their behavior and interactions. By observing runtime behavior, dynamic analysis provides more detailed information. Dynamic analysis methods are more robust as they can directly observe the actual behaviors of malware. 

With the development of deep learning, leveraging deep learning models to analyze malware behavior has become a significant detection approach \cite{8821336,shaukat2022novel}. Since deep learning-based models have shown strong sequential pattern learning capabilities \cite{chai2022dynamic}, the behaviors of malware can be effectively utilized for modeling the sequential features to identify anomalies \cite{gopinath2023comprehensive}. Several studies have demonstrated the effectiveness of incorporating syscalls for malware detection \cite{liu2020statistical,soni2019behavioral,kishore2022applying}. By analyzing the syscall sequences, these models can learn patterns and correlations indicative of malicious behavior. They can distinguish between benign software and malware based on the distinctive syscall patterns associated with each category. Similarly, the analysis of API call sequences is another valuable approach in the field of malware detection \cite{hardy2016dl4md,rhode2018early,natani2013malware}. These approaches achieve detection by learning  statistical features \cite{natani2013malware} or sequentical features \cite{hardy2016dl4md,rhode2018early}, and have more accurate identification of unknown malware. 

\subsection{Adversarial Attacks on Deep Learning Models}

Adversarial attacks on deep learning models have been a prominent research topic in recent years. Szegedy et al. \cite{szegedy2013intriguing} first revealed the vulnerabilities of deep learning models to adversarial attacks. 
Since then, various adversarial attack methods \cite{goodfellow2014explaining,kurakin2016adversarial, carlini2017towards} continue to emerge.
The majority of these methods primarily concentrate on perturbing image inputs to deceive deep learning models in the field of computer vision. Recently, the investigation of adversarial attacks on sequence-based classifiers has emerged as a hot topic. In the domain of malware detection, adversarial attacks against sequence-based classifiers have been explored \cite{papernot2017practical,rosenberg2020query}. These attacks focus on adding API calls that do not affect the functionality of the malware or only produce unrelated effects, making it harder for classifiers to detect them.

Recent work called (LAM) \cite{herath2021real}, has been developed to deceive sequence-based anomaly detectors. LAM applies reinforcement learning to modify behavior sequences and has proven to be effective in adversarial attacks.
Another notable work by Siyang Lu et al. \cite{lu2023black} proposed two methods to attack deep neural networks used for log anomaly detection: an attention-based attacker and a gradient-based attacker. Both methods identify susceptible behaviors within the sequence and make effective modifications to them.
These methods directly manipulate data, and the modified sequence is challenging for guiding program modifications.

% \subsection{Defense Mechanisms Against Adversarial Attacks}

% Defending against adversarial attacks is an important research direction, and several defense methods have been proposed. Adversarial training \cite{szegedy2013intriguing} is a common approach that adds adversarial samples to the training set, allowing the classifier to learn the distribution of adversarial samples and improve its robustness. Another strategy is defensive distillation \cite{papernot2016distillation}, which trains the model to produce probabilities over classes rather than hard predictions, making it more robust to adversarial examples. 
% Model ensemble techniques, such as DroidEvolver \cite{xu2019droidevolver}, build a model pool composed of multiple machine learning models and perform classification through multi-model voting, which enhances the overall robustness against adversarial attacks.

\section{Discussion} \label{s:Discussion}

% \subsection{Defense against the Attack}

\subsection{Limitations and Future Works}
Our approach might face some potential challenges. First, our approach may need continual updates or adaptations to keep pace with rapidly evolving malware tactics. In future work, we plan to incorporate continuous learning techniques to solve this challenge. Furthermore, the performance of our approach might vary when applied to other advanced classifiers. Future work could explore more detection models to mitigate this limitation.

\subsection{Defense and Responsible Disclosure}

% \textbf{Defense.} 
Our approach involves adding benign behaviors to the original sequence to generate adversarial samples. A possible mitigation strategy could be to extract critical behavioral sub-sequences during the data preprocessing stage, thereby filtering out irrelevant behaviors. Another alternative approach could involve the development of models that can adaptively identify various key patterns of malicious sub-sequences. This would allow for the exclusion of irrelevant sequence behaviors that might be introduced in the adversarial sequences.

% We place a high emphasis on ethics and conduct conceptual validation using publicly available datasets first. Subsequently, we modify malicious software on our own operating systems for the purpose of feasibility verification. We never intend to exploit any vulnerabilities or cause harm on other users' systems.
\textbf{Responsible Disclosure.} 
We have already established contact with several security companies involved in malware detection or log analysis. We've alerted them to the potential risks arising from anomaly detection based on serialized data, specifically that adversarial sample data may bypass their detection products. For instance, Antiy \cite{Antiy} has acknowledged our findings, engaged in discussions about vulnerabilities in the detection models, and is searching for similar issues while making corresponding modifications in their products. We will continue to maintain contact with other security teams and collaborate to enhance the adversarial robustness of malware detection based on serialized data.

\section{Conclusion} \label{s:conclusion}
In this paper, we have presented a novel adversarial attack method for behavior sequence-based classifiers, leveraging deep reinforcement learning to generate adversarial samples while preserving the functionality of malware. Our approach demonstrated effectiveness in evading various target models, outperforming state-of-the-art baselines in our experiments. The practicality assessment also indicated that our approach is computationally efficient, adaptable, and highly applicable in real-world settings. This work highlights the need for developing more effective defense mechanisms and encourages further research on enhancing the security of behavior sequence-based classifiers.

% In light of these limitations, we plan to study the following measures, including the development of adaptive algorithms to handle evolving malware behaviors, investigation into scalability solutions, and thorough validation in diverse real-world scenarios.

% or


%\appendices

%\section{}

\ifCLASSOPTIONcompsoc
  \section*{Acknowledgments}
\else
  \section*{Acknowledgment}
\fi

This work was supported by the National Key R\&D Program of China (No. 2021YFB2012402), the National Natural Science Foundation of China under Grants No. 62302122 and No. 62172123, and the Natural Science Foundation of Heilongjiang Province of China under Grants No. LH2023F017.

\ifCLASSOPTIONcaptionsoff
  \newpage
\fi



%\bibliographystyle{IEEEtran}
%\begin{thebibliography}{1}


%\end{thebibliography}

\bibliographystyle{IEEEtran}
\bibliography{sec}


\begin{IEEEbiography}[{\includegraphics[width=1in,height=1.25in,clip,keepaspectratio]{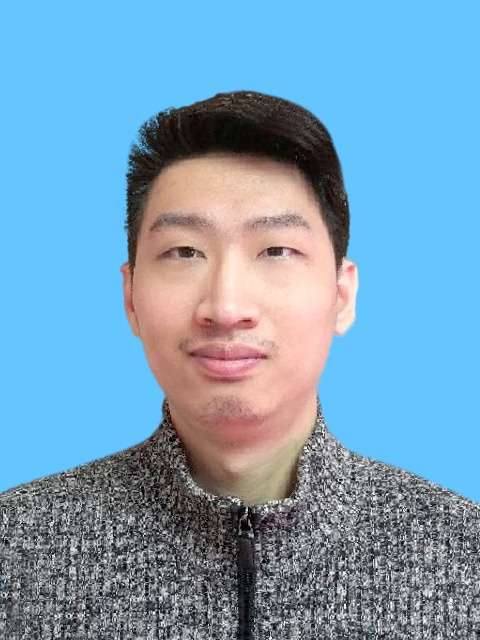}}]{Kai Tan} is a PhD candidate from the Harbin Institute of Technology, China. His research focuses on cloud security.
\end{IEEEbiography}

\begin{IEEEbiography}[{\includegraphics[width=1in,height=1.25in,clip,keepaspectratio]{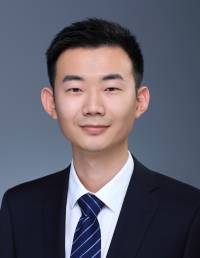}}]{Dongyang Zhan} is an assistant professor in School of Cyberspace Science at Harbin Institute of Technology. He received the B.S. degree in Computer Science from Harbin Institute of Technology from 2010 to 2014. From 2015 to 2019, he has been working as a Ph.D. candidate in School of Computer Science and Technology at HIT. His research interests include cloud computing and security.
\end{IEEEbiography}

\begin{IEEEbiography}[{\includegraphics[width=1in,height=1.25in,clip,keepaspectratio]{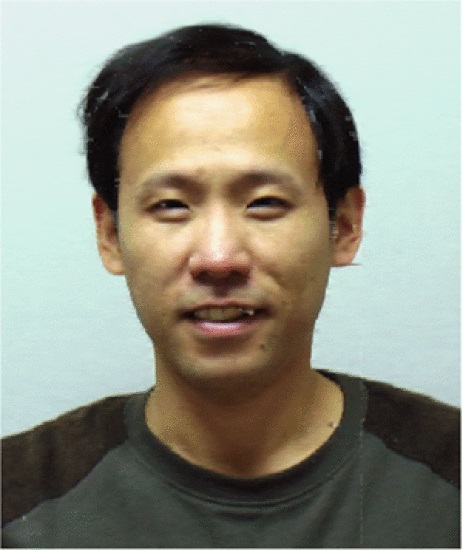}}]{Lin Ye}
received the Ph.D. degree at Harbin Institute of Technology in 2011. From January 2016 to January 2017, he was a visiting scholar in the Department of Computer and Information Sciences, Temple University, USA. His current research interests include network security, peer-to-peer network, network measurement and cloud computing.
\end{IEEEbiography}

\begin{IEEEbiography}[{\includegraphics[width=1in,height=1.25in,clip,keepaspectratio]{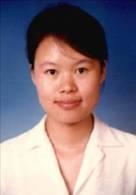}}]{Hongli Zhang}
received her BS degree in Computer Science from Sichuan University, Chengdu, China in 1994, and her Ph.D. degree in Computer Science from Harbin Institute of Technology (HIT), Harbin, China in 1999. She is currently a Professor in School of Cyberspace Science in HIT. Her research interests include network and information security, network measurement and modeling, and parallel processing.
\end{IEEEbiography}

%\newpage

\begin{IEEEbiography}[{\includegraphics[width=1in,height=1.25in,clip,keepaspectratio]{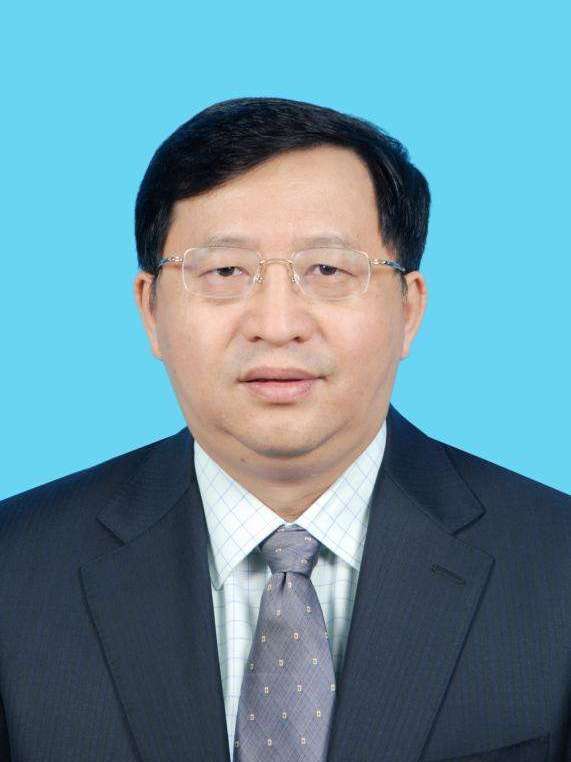}}]{Binxing Fang} (Member, IEEE) received the M.S. degree in computer science and technology from Tsinghua University, Beijing, China, in 1984, and the Ph.D. degree in computer science and technology from the Harbin Institute of Technology, Harbin, China, in 1989. He is currently a Professor with the Department of Computer Science and Technology, Harbin Institute of Technology (Shenzhen), Shenzhen, China. He is also a member of the Chinese Academy of Engineering, Beijing. His current research interests include computer networks, information and network security, and artificial intelligence security.
\end{IEEEbiography}


%\vfill


\end{document}